\documentclass[11pt,preprint]{aastex}
\usepackage{emulateapj5}
\usepackage{apjfonts}

\newcommand{\myemail}{chenlh98g@mails.tsinghua.edu.cn}

\def\BHM{M_{\rm BH}}
\def\dotm{\dot{m}}
\def\DOTM{\dot{M}}
\def\sax{\em BeppoSAX}
\def\asca{\em ASCA}

\def\chandra{\em Chandra}
\def\rxte{\em RXTE}
\def\sunm{M_{\odot}}

\shorttitle{Slim Disks with Transition Regions }

\shortauthors{Chen \& Wang}

\begin{document}

\title{Slim Disks with Transition Regions and Applications to Microquasars \\
and Narrow Line Seyfert 1 Galaxies}

\author{Lin-Hong Chen}
\affil{Center for Astrophysics, Tsinghua University, Beijing 100084, China}
\email{\myemail}

\author{Jian-Min Wang}
\affil{Laboratory of High Energy Astrophysics, Institute of High Energy Physics, Chinese Academy 
of Sciences,\\ Beijing 100039, China}

\slugcomment{to Accepted for Publication by {\em The Astrophysical Journal}}
\shorttitle{Slim Disk and Application to Micro-quasars and Narrow Line Seyfert 1 Galaxies}
\shortauthors{CHEN \& WANG}

\begin{abstract}
Slim disks have been received much attention because of the increasing evidence for the
super-critical accreting objects. In this paper, we make an attempt to construct a
unified model, in which the viscosity and the dimensionless accretion rate can span
rather wide ranges. We replace blackbody radiation under diffusion approximation with
a bridged formula, which accounts for both blackbody radiation and thermal bremsstrahlung
in optically-thick and -thin cases, respectively. Thus this allows us to investigate
the structures of and the emergent spectra from slim disks in a wider parameter space,
covering transition regions from optically thick to optically thin.
We show that there is a maximum transition radius, roughly $R_{\rm tr, max}/R_g\sim 50$
when $\dot{M}/\dot{M}_{\rm C}\sim 15$. The emergent spectra from the unified model of
the accretion disk have been calculated. A simple model of hot corona above the slim disk
is taken into account for the hard X-ray spectrum in this paper based on Wang \& Netzer (2003). 
We have applied the present model to the microquasar GRS 1915+105, narrow line Seyfert 1 galaxies
RE J1034+396 and Akn 564. Our model can explain well the broadband X-ray spectra
of narrow line Seyfert 1 galaxies, microquasars and possible ultra-luminous compact X-ray sources.
The present model can be widely applied to the candidates of super-critical accreting objects.
\end{abstract}


\section{Introduction}
It is believed that accretion disks around black holes are powering in many kinds of
celestial objects. The most important scale in the description of accretion disks
is the critical accretion rate, defined by
$\dot{M}_{\rm C}=L_{\rm Edd}/\eta c^2=2.6\times 10^{18}M_{\rm BH}/\sunm$ g/s,
where the accretion efficiency $\eta=1/16$ is for Schwarzschild black holes whose gravity
effects are approximated by the pseudo-Newtonian potentials. The structures of accretion disks
rely greatly on the mass of the black hole ($\BHM$), the accretion rate ($\DOTM$)
and the viscosity ($\alpha$), and hence the emergent spectrum from the disk.
According to the dimensionless accretion rate $\dotm=\dot{M}/\dot{M}_{\rm C}$, 
the disk structures can be classified:
(1) optically thin advection-dominated accretion flows (ADAFs; Narayan \& Yi 1994),
advection-dominated inflow-outflows (ADIOs; Blandford \& Begelman 1999) and
convection-dominated accretion flows (CDAFs; Narayan et al. 2000; Quataert \& Gruzinov 2000)
when $\dotm<\alpha^2$, (2) optically thick and geometrically thin disks (Shakura \& Sunyaev 1973)
if $\alpha^2<\dotm<0.2$, (3) slim disks (Muchotrzeb \& Paczy\'{n}ski 1982; Abramowicz et al. 1988;
Chen \& Taam 1993) once $\alpha^2<\dotm<100$. Chen et al. (1995) presented a unified model,
but they {\it assumed} a Keplerian angular momentum distribution in the disk. The unified model
to cover all these regimes has not been sufficiently understood, especially we still
lack the emergent spectrum of such a model for the comparison with observations. 

There is growing evidence from current observations that the accretion rates of a significant
fraction of celestial objects span quite wide ranges from very low to super-critical in X-ray 
binaries, microquasars and active galactic nuclei (AGNs). In our Galaxy, some ultra-luminous
compact X-ray sources (ULXs), microquasars such as GRS 1915+105 and SS 433, show large bolometric
luminosities and high color temperatures (>1 keV) in their high states (Makishima et al. 2000;
Watarai et al. 2001; Ebisawa et al. 2003). This provides strong evidence for the appearance of
slim disks in these objects. The Eddington ratios ($L/L_{\rm Edd}$) in AGNs and quasars
have been reliably estimated for revealing the structures of the disks in AGNs and quasars,
since the black hole masses can be estimated by the reverberation mapping relation
(Kaspi et al. 2000) and the relation with their host galaxies (e.g. McLure \& Dunlop 2001).
It has been suggested that some of AGNs have close or super critical accretion rates 
(Collin et al. 2002; Netzer 2003; Vestergaard 2003; Wang 2003; Szuszkiewicz 2003),
whose luminosities seemingly exceed $L_\mathrm{Edd}$ (if the estimations of black hole masses
are reliable). Collin et al. (2002) have shown that about half of the 34 AGNs,
of which the masses are determined by reverberation mapping, have super-critical accretion rates.
Vestergaard (2003) extended the empirical relation of reverberation mapping in the sample
of Kaspi et al. (2000) to the cases of high redshit quasars (see also McLure \& Jarvis 2002).
She found there are a significant number of quasars with super-Eddington luminosities,
in either intermediate redshift range $1.5\leq{z}\leq{3.5}$ or high redshift range $z>3.5$.
Willott et al. (2003) found that the quasar SDSS J1148+5251 at $z=6.41$ is radiating at
the Eddington luminosity with the estimated mass $\BHM=3\times10^9M_\odot$.
Wang (2003) has also discovered a handful of the candidate super-critical accretors,
using the limit relation between the black hole mass and FWHM(H$\beta$),
$M_{\rm BH}=(2.9-12.6)\times10^6\left[\upsilon_{\rm FWHM}/10^3\right]^{6.67}M_\odot$.
This relation is a result by combining the empirical reverberation mapping and the features
of the emergent spectra from slim disks. Although the spectra of 110 bright AGNs in the sample
of Laor (1990) are fitted well by standard thin disks (Shakura \& Sunyaev 1973),
those super-Eddington objects can never be explained within the framework of the standard
disk theory. A wide range of dimensionless accretion rates should exist in various kinds of
celestial objects, but there is not a unified model of accretion disks to describe
the varieties of accretion.

\figurenum{1}
\centerline{\includegraphics[angle=-90.0,width=8.5cm]{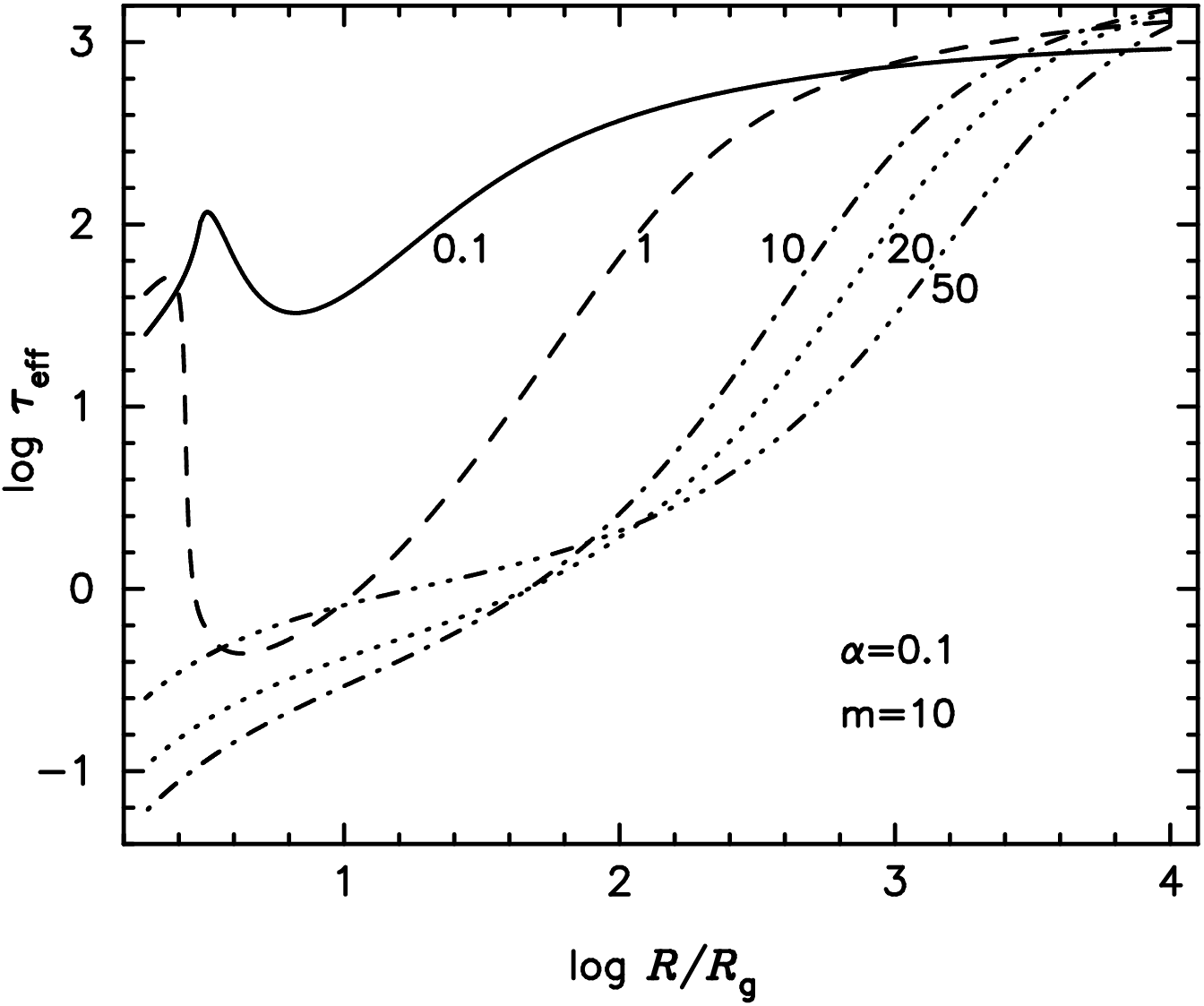}}
\figcaption{Effective optical depth ($\tau_\mathrm{eff}$) of the classical slim disk (see \S2.1).
The corresponding accretion rate is labeled. The effective optically thin region
($\tau_\mathrm{eff}\le1$) should be noted for $1<\dotm\leq{50}$.} 
\label{taueff}
\vglue 0.2cm

The significance of slim disks has been exhibited when they are applied to narrow line
Seyfert 1 galaxies (NLS1s; Wang et al. 1999; Mineshige et al. 2000; Wang \& Netzer 2003;
Kawaguchi 2003), Galactic black hole candidates (Watarai et al. 2000, Fukue 2000) and ULXs
(Watarai et al. 2001) for the strong soft X-ray emission and the violent variability.
However, slim disk models in above literatures suffer from the problem that the effective optical
depth is smaller than unity in the inner disk region. The slim disks have been thought to work
in these objects with high accretion rates, but the emergent spectra from the disks are not
satisfactory enough. We list the models and spectra studied by far in Table 1, in which
the main assumptions and features in these models are also given.
The emergent spectra from the accretion disks of a wide range of $\dot{m}$ remain open so far,
especially only Wang \& Netzer (2003) calculated the spectrum from the slim disk with hot corona
based on the self-similar solution. We noted that most of the models assumed a low viscosity
so that the transition to an optically-thin region does not appear in their models (see equation 1). 
Szuszkiewicz et al. (1996) calculated the emergent spectrum with the modified blackbody radiation
and explained the soft X-ray excesses claimed for some AGNs with slim disks. The vertical structures
and the continuum spectra have been computed by Wang et al. (1999) and Shimura \& Manmoto (2003). 
Szuszkiewicz et al. (1996) presented the optically thin region with a small viscosity parameter
$\alpha=0.001$ and the accretion rates $1<\dotm<50$. Kawaguchi (2003) also pointed out these regions
in the accretion disks around super massive black holes (SMBHs). But they did not provide
a reasonable treatment in the optically-thin hot region. In this paper, we show that a
much wider region, where the effective optical depth $\tau_\mathrm{eff}\le1$, appears when the
viscosity increases to $\alpha=0.1$ for a stellar-mass black hole accretion (Fig. 1).
It can be seen that the transition radii from optically thick to thin locate at least several
decades $R_\mathrm{g}$ ($R_\mathrm{g}=2GM/c^2$, the Schwarzschild radius)
for the accretion rates $10\le\dot{m}\le50$. In such large inner regions,
the diffusion approximation for radiative transfer in the slim disks (Abramowicz et al. 1988)
breaks down. We can not avoid the calculations of the emergent spectra from such disks
having transition regions and hot corona, if connect the slim disks with observations.

\begin{center}
\footnotesize
{\sc Table 1 \\}
{\sc A Brief Summary: Structures and Spectra of Slim Disk Models}
\vglue 0.2cm
\begin{tabular}{lcccccl}\hline \hline
Ref.    & Model        &$\alpha$ & vertical & $\tau_{\rm eff}$  & Corona & Spectrum \\ \hline
MP82    & global       & low     & no       & $\gg 1$           & no     & no      \\
A88     & global       & low     & no       & $\gg 1$           & no     & no      \\
CT93    & global       & low     & no       & $\gg 1$           & no     & no      \\
C95     & Kepler       & $\sim$  & no       & $\sim $           & no     & no      \\
H96     & global       & $\sim$  & no       & $\sim $           & no     & no      \\
SMA96   & global       & low     & no       & $\gg 1$           & no     & yes$^1$ \\
B98     & global       & $\sim$  & no       & $\sim $           & no     & no      \\
W99     & global       & low     & yes      & $\gg 1$           & no     & yes$^2$ \\
WZ99    & self-similar & low     & no       & $\gg 1$           & no     & no      \\
F00     & self-similar & low     & no       & $\gg 1$           & no     & yes$^3$ \\
M00     & global       & low     & no       & $\gg 1$           & no     & yes$^3$ \\
WFT00   & global       & low     & no       & $\gg 1$           & no     & yes$^3$ \\
A01     & global       & low     & no       & $\gg 1$           & no     & no      \\
ZTS01   & global       & $\sim$  & yes      & $\sim $           & no     & yes$^2$ \\
K03     & global       & low     & no       & $\gg 1$           & no     & yes$^4$ \\
SM03    & global       & low     & yes      & $\gg 1$           & no     & yes$^2$ \\
WN03    & self-similar & low     & no       & $\gg 1$           & yes    & yes$^4$ \\
present & global       & $\sim$  & no       & $\sim $           & yes    & yes$^4$ \\ \hline
\end{tabular}
\parbox{3.2in}
{\vglue 0.1cm
\noindent\footnotesize
{\sc Notes.}--- \\
1: modified blackbody; 2: solving radiative transfer equation; 
3: local blackbody approximation; 4: modified blackbody including Comptonization. \\
\vglue 0.01cm
\noindent{\sc Reference.}--- \\
A88: Abramowicz et al. (1988); A01: Artemova et al. (2001); B98: Beloborodov (1998); 
CT93: Chen \& Taam (1993); C95: Chen et al. (1995); F00: Fukue (2000); H96: Honma (1996);
K03: Kawaguchi (2003); M00: Mineshige et al. (2000); MP82: Muchotrzeb \& Paczy\'{n}ski (1982); 
SM03: Shimura \& Manmoto (2003); SMA96: Szuszkiewicz, Malkan, \& Abramowicz (1996); 
W99: Wang et al. (1999); WZ99: Wang \& Zhou (1999);
WFT00: Watarai, Fukue, \& Takeuchi (2000); WN03: Wang \& Netzer (2003);
ZTS01: Zampieri, Turolla, \& Szuszkiewicz (2001)
}
\end{center}

This paper is organised by following: the basic formulations for the calculations of structures and
spectra are given in \S2, and the numerical results are presented in \S3. We find that the transonic
location changes due to the presence of transition region. So does the structure in the inner disk
region. We pay our attentions on how the emergent spectrum gradually changes with the accretion rate
increasing from sub- to super-critical. The transition to an optically-thin region modifies
the emergent spectrum, especially in the EUV and soft X-ray bands. We apply the present slim disk
model with transition region to Galactic black hole candidates, one microquasar, and two NLS1s sources
in \S4. The conclusions are laid in the final section.

\section{Basic Model}
According to the model of Shakura \& Sunyaev (1973), the effective optical depth will be less
than unity at the radius
\begin{equation}
R_{\rm tr}/R_\mathrm{g} \approx 25\alpha^{34/93}\dot{m}^{64/93}m^{2/93},
\end{equation}
where we have neglected the factor $(1-r^{-1/2})$ ($r=R/R_\mathrm{g}$ and $m=M/M_\odot$).
This formula shows that the transition covers the region where most of the gravitational energy
is released. Within this region the radiation diffusion approximation does not work. Additionally,
the transonic point is also located in this region. The transition radius $R_{\rm tr}$ is
very sensitive to the accretion rate and the viscosity. We will focus on the global structure
and the emergent spectrum from the slim disk with transition region. The equations are described
as below.

\subsection{Equations of slim disks with transition regions}
The basic equations of slim disks are taken from Muchotrzeb \& Paczy\'{n}ski (1982) and Abramowicz
et al. (1988), but we deal with the radiative cooling more carefully. The vertical static equilibrium
is assumed, namely $H=B_2c_\mathrm{s}/\Omega_\mathrm{K}$, where $H$ is the half-height of the disk
at the radius $R$, $c_\mathrm{s}=\sqrt{P/\rho}$ the sound speed, 
$\Omega_\mathrm{K}=\sqrt{GM/R^3}/(1-R_\mathrm{g}/R)$ the Keplerian angular velocity, $P$ the total
pressure and $\rho$ the density. Unless stated, all physical quantities are referred to their
equatorial plane values. We use the Shakura-Sunyaev viscosity prescription, 
$\tau_{r\varphi}=-\alpha{P}$,
where $\tau_{r\varphi}$ is the viscous torque tensor and $\alpha$ is a phenomenological viscosity
parameter. The pseudo-Newtonian potential (Paczy\'{n}ski \& Witta 1980) is adopted in our
calculations for Schwartzschild black holes.

The four equations of mass, radial momentum, angular momentum and energy conservations control the
structure of the slim disk. The momentum equation in the radial direction is
\begin{equation}
\frac{1}{\rho}\frac{dP}{dR}-(\Omega^2-\Omega_\mathrm{K}^2)R+\upsilon_\mathrm{R}
\frac{d\upsilon_\mathrm{R}}{dR}=0,
\end{equation}
where $\upsilon_\mathrm{R}$ is the radial drift velocity of the accreting gas, $\Omega$ the angular
velocity. The specific angular momentum and the Keplerian angular momentum are defined by $l$ and
$l_\mathrm{K}=\Omega_\mathrm{K}R^2$. The angular momentum conservation reads 
\begin{equation}
\dot{M}(l-l_\mathrm{in})=4\pi{R^2}H\alpha{P},
\end{equation}
where $l_\mathrm{in}$ is the eigenvalue of angular momentum at the inner disk edge. The mass
accretion rate, $\dot{M}$, is given by
\begin{equation}
\dot{M}=B_52\pi{R}\Sigma{\upsilon_\mathrm{R}},
\end{equation}
where $\Sigma=2\rho{H}$ is the surface density. 

In the classical slim disk model of Abramowicz et al. (1988),
the viscosity parameter $\alpha$ is very low. This ensures the validity of the radiation
diffusion approximation for the radiative cooling. However, the low viscosity only covers
a very small parameter space. When the viscosity increases, the density decreases
so that the effective optical depth will be less than unity. In such a case the radiation
diffusion approximation breaks down. In the regime of slim disk,
we adopt a bridged formula 
\begin{equation}
%Q_\mathrm{rad}=\frac{4\sigma\/T^4}{3\tau_\mathrm{R}} \frac{c_0(\mathrm{e}^{\tau_\mathrm{eff}}-1)
%               \tau_\mathrm{eff}}{1+c_0(\mathrm{e}^{\tau_\mathrm{eff}}-1)\tau_\mathrm{eff}}A
Q_\mathrm{rad}=B_3\frac{4\sigma\/T^4}{3(\tau_\mathrm{es}+\tau_\mathrm{abs})}
\frac{(1-\mathrm{e}^{-\tau_{\rm eff}})A\tau_{\rm eff}}
     {\mathrm{e}^{-\tau_{\rm eff}}+(1-\mathrm{e}^{-\tau_{\rm eff}})A\tau_{\rm eff}},
\end{equation} 
for the vertical radiation transport (see also Wandel \& Liang 1991; Chen et al. 1995). Here $\sigma$
is the Stefan-Boltzmann constant, $T$ the mid-plane temperature, and the effective optical depth
$\tau_\mathrm{eff}=\sqrt{(\tau_\mathrm{es}+\tau_\mathrm{abs})\tau_\mathrm{abs}}$, where we take
the electron scattering depth $\tau_\mathrm{es}=0.34\rho H$, the absorption depth
$\tau_\mathrm{abs}=6.4\times10^{22}\rho^2T^{-3.5}H$ for stellar-mass black holes, and 
$\tau_\mathrm{abs}=1.92\times10^{23}\rho^2T^{-3.5}H$ for SMBHs due to bound-free processes
(Czerny \& Elvis 1987). This bridged formula works for a transition from optically thick
to optically thin. The Comptonization amplification factor $A$ is adopted from
Wandel \& Liang (1991) for the simplicity. Compared with another type of bridged formula
(Szuszkiewicz \& Miller 1998; Narayan \& Yi 1995), this bridged formula and the definition of
$\tau_\mathrm{eff}$ are consistent with those in spectral calculations described in below subsection.
Heating due to viscosity, $Q_\mathrm{vis}$, is cooled by local radiation,
$Q_\mathrm{rad}$, and global radial advection, $Q_\mathrm{adv}$, so the energy
equation can be symbolically written as  
\begin{equation}
Q_\mathrm{vis}=Q_\mathrm{rad}+Q_\mathrm{adv},
\end{equation}
where $Q_\mathrm{adv}=B_1\frac{\dot{M}}{4\pi{R^2}}\frac{P}{\rho}\xi$, 
$\xi=-(12-10.5\beta)d\ln{T}/d\ln{R}+(4-3\beta)d\ln\rho/d\ln{R}$ is the dimensionless advection factor, 
$\beta$ is the ratio of gas to total pressure. The radial energy transport is neglected.
The state equation is
\begin{equation}
P=\frac{k_B}{\mu{m_{\rm H}}}\rho{T}+\frac{Q_{\rm rad}(\tau_{\rm es}+\tau_{\rm abs})}{c},
\end{equation}
where $\mu=0.617$ is the mean molecular weight, $k_B$ the Boltzmann constant, $m_{\rm H}$
the mass of hydrogen. In this paper, we adopt $B_1=0.67$, $B_2^2=B_3=B_4=6$ and $B_5=0.5$,
just as Paczy\'{n}ski \& Bisnovatyi-Kogan (1981); Muchotrzeb \& Paczy\'{n}ski (1982); Muchotrzeb (1983).

Two-temperature plasma is used in Wandel \& Liang (1991), we assume the accreting gas is
single-temperature over the entire disk, i.e. the energy transfer between the electrons and protons
is very efficient. This assumption can be justified by comparing the timescales of free-free cooling
and of Coulomb interaction between electrons and protons. The free-free cooling timescale is
$t_{\rm ff}=2\times10^2\rho_{10}^{-2}T_6^{1.5}$ seconds and the timescale of the Coulomb interaction
between electrons and protons is $t_{\rm Coul}\approx2\times10^{-4}\rho_{10}^{-1}T_6^{1.5}$ seconds,
where $\rho_{10}=\rho/10^{-10}$ and $T_6=T/10^6$. When $t_{\rm Coul}\ge t_{\rm ff}$, i.e.
$\rho\le10^{-16}$ $\mathrm{g}/\mathrm{cm}^3$, the plasma has two temperatures. We find that all
the calculated $\rho$ in this paper are consistent with the single-temperature assumption. 

\subsection{Formulations for emergent spectra}
We employ the method described by Czerny \& Elvis (1987) to calculate the emergent spectra from
accretion disks, considering the effect of opacity due to elastic electron scatterings
and the effect of energy exchange in nonelastic scatterings (i.e. Comptonization).
The photons are scattered before they escape from the surface of the disk (Rybicki \& Lightman 1979),
the emergent spectra are thus modified. However, Laor \& Netzer (1989) pointed out that
Comptonization is not important for disks around $10^9$ $M_\odot$ black holes and $L/L_{\rm Edd}<0.3$.
In slim disks, the transition region appears, where the free-free absorption is not important
but the scattering becomes essential. Wang \& Netzer (2003) have shown this process is very
important in slim disks, as confirmed by Kawaguchi (2003), see also \S3.3 in this paper.

The local radiation intensity 
\begin{equation}
I_\nu(r)=B_\nu(T_\mathrm{s})f_\nu(T_\mathrm{s},r),
\end{equation}
where $B_\nu$ is the Planck function, $T_\mathrm{s}$ the temperature on disk surface, and $f_\nu$
describing the departure from blackbody radiation including Comptonization is
\begin{equation}
f_\nu(T_\mathrm{s},r)=f_\mathrm{ff}(1-f_\mathrm{th})+C.
\end{equation}
Here $f_\mathrm{ff}$ and the fraction of thermalized photons $f_\mathrm{th}$ are given as
\begin{equation}
f_\mathrm{ff}(T_\mathrm{s},r)=\frac{2(1-\mathrm{e}^{-2\tau_\mathrm{eff}})}
                              {1+\sqrt{1+\tau_\mathrm{es}/\tau_\mathrm{abs,\nu}}},
\end{equation}
and
\begin{equation}
f_\mathrm{th}(T_\mathrm{s},r)=\mathrm{exp}\left\{-\frac{\ln(k_BT_\mathrm{s}/h\nu)}{\tau_{\nu*}^2
                              \ln[1+\Xi_e +16\Xi_e^2]}\right\},
\end{equation}
respectively, where the electron's dimensionless temperature
$\Xi_e=4k_{B}T_{\rm s}/m_ec^2$, and
$\tau_{\nu*}=(\tau_\mathrm{es}+\tau_\mathrm{abs,\nu})/(1+\tau_\mathrm{eff})$. $f_\mathrm{ff}$
restores the modified blackbody radiation for $\tau_\mathrm{eff}\gg1$, and describes
the thermal bremsstrahlung from an optically thin medium for low $\tau_\mathrm{eff}$.
The Comptonization normalization constant is calculated by
\begin{equation}
C=\frac{3k_BT_\mathrm{s}\int_0^\infty\/f_\mathrm{th}\pi\/I_\nu/h\nu\/d\nu}{\sigma\/T_\mathrm{s}^4}.
\end{equation}
The surface temperature $T_\mathrm{s}$ is reached by the balance between the radiated energy and the
dissipated gravitational energy
\begin{equation}
Q_\mathrm{rad}=\int_0^\infty\pi\/I_\nu(R,T_{\rm s})\/d\nu
\end{equation}
This simple method for $T_{\rm s}$ allows us to avoid the calculations of the vertical structure
of the slim disk, and the accuracy of the spectrum is enough for the present aims. The total
emergent spectrum is derived by integrating $I_\nu(r)$ over the radius,
\begin{equation}
L_\nu=2\int_{R_\mathrm{in}}^{R_\mathrm{out}}\pi\/I_\nu(R)2\pi\/RdR.
\end{equation}

The hard X-ray emission from the hot corona above the slim disk is also calculated based on the
detailed description in Wang \& Netzer (2003). The most important characterized parameter $f$,
the fraction of the gravitational energy released in the
hot corona, is assumed in our calculation. However, this factor $f$ is confined by the fact that
the hot corona should keep, at least, as optically thin medium, otherwise it will be rapidly
cooled. This limit can be roughly estimated by $\tau_{\rm es}<1$ at the last stable radius with
the assumption that the radial velocity is light speed. Wang \& Netzer (2003) showed 
\begin{equation}
f<0.0375\left(\frac{\dot{m}}{10}\right)^{-1}.
\end{equation} 
This limit shows that the hot corona becomes weak with the increase of the accretion rate.
Indeed the hard X-ray spectra from microquasars and NLS1s show such a property. 
In our calculations, we thus neglect the feedback of hot corona to the slim disk.
This is justified because the factor $f$ is small for a high accretion rate.
However, a detailed treatment including this feedback for the case of intermediate
accretion rate will be carried out in a separate paper.

\section{Numerical Results}
The equations (2-7) can be reduced to two first-order differential equations about two variables,
$\rho$ and $T$. We use the shooting method to solve these equations with appropriate
boundary conditions. We assume an outer boundary at $10^4$ $R_{\rm g}$, where Shakura-Sunyaev
solutions (the gas pressure-dominated and $\kappa_{\rm abs}\gg\kappa_{\rm es}$) are available.
At this boundary, we firstly choose the derivatives of $\rho$ and $T$ obtained from Shakura-Sunyaev
disk solutions, substitute them into the two differential equations, then calculate the values
of $\rho$ and $T$ by iterations. Although they are very close to Shakura-Sunyaev values,
the new values of $\rho$ and $T$ allows the inward integration to run on the right way.
The general relativistic effects are simulated by the pseudo-Newtonian potential near the
horizon of black holes. Fourth-order Runge-Kutta method is employed to integrate the two
differential equations to the inner disk edge $R_\mathrm{in}$, where $R_\mathrm{in}$ is
taken as a free parameter as other authors did (e.g. Mineshige et al. 2000;
Wang \& Netzer 2003). We use the torque-free condition at the inner boundary. 
The shooting method determines the eigen angular momentum $l_\mathrm{in}$, provided the solution
satisfies the regularity condition at the critical point close to the transonic radius.

After we obtain the global transonic solutions, the basic physical quantities in the accretion
flow and its emergent spectrum are given for stellar black holes ($10M_\odot$) and SMBHs
($10^6M_\odot$) in this section. We show in detail how the transition to an optically thin region
appears with the change of accretion rate. 

\subsection{Transonic solutions}
\figurenum{2}
\centerline{\includegraphics[angle=-90,width=8.5cm]{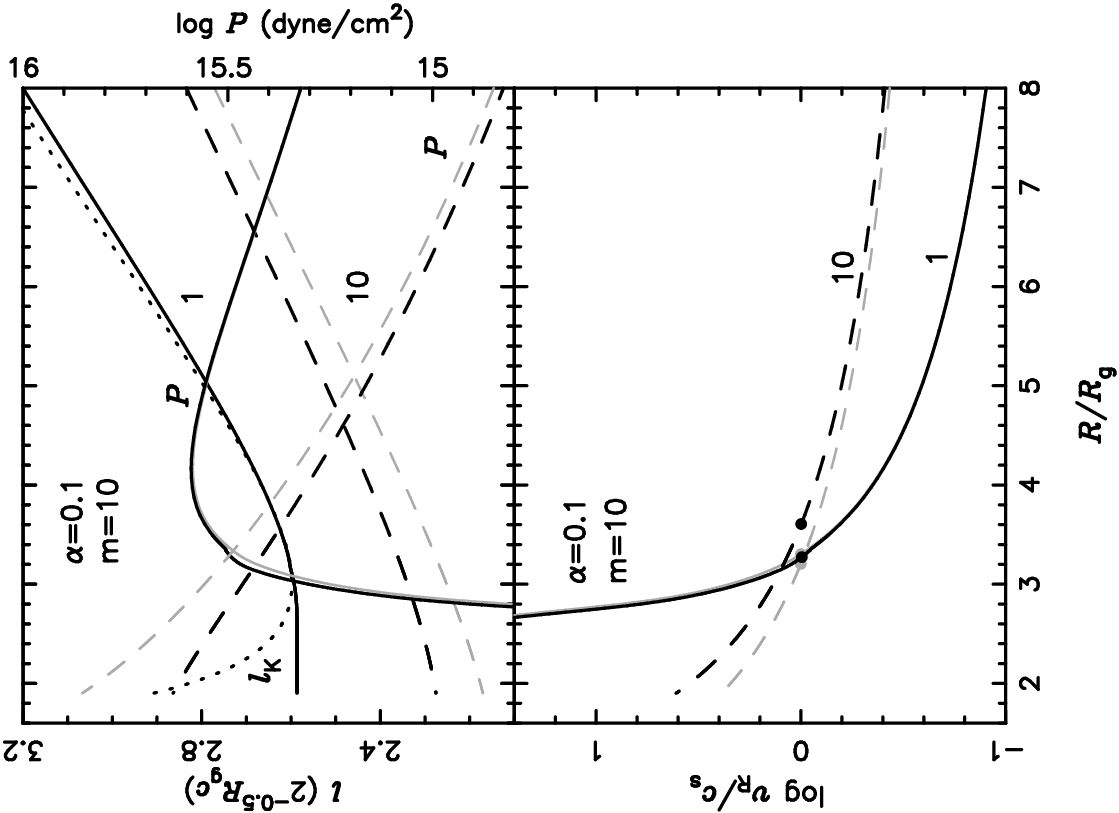}}
\figcaption{Angular momentum ($l$) and total pressure ($P$; top panel), ratio of radial
velocity to sound speed ($\upsilon_\mathrm{R}/c_\mathrm{s}$; bottom panel) for $\dotm=1$
(solid lines) and $\dotm=10$ (dashed lines). The total pressure of $\dot{m}=1$ is shifted
upward by one logarithmic unit. The transonic locations are marked by solid points.
The grey lines are the results of the classical slim disks calculated in this paper
(the same for Figs. 3-5, 9, 10 and 12).} 
\label{transonic}
\vglue 0.2cm

\figurenum{3}
\centerline{\includegraphics[angle=-90.0,width=8.5cm]{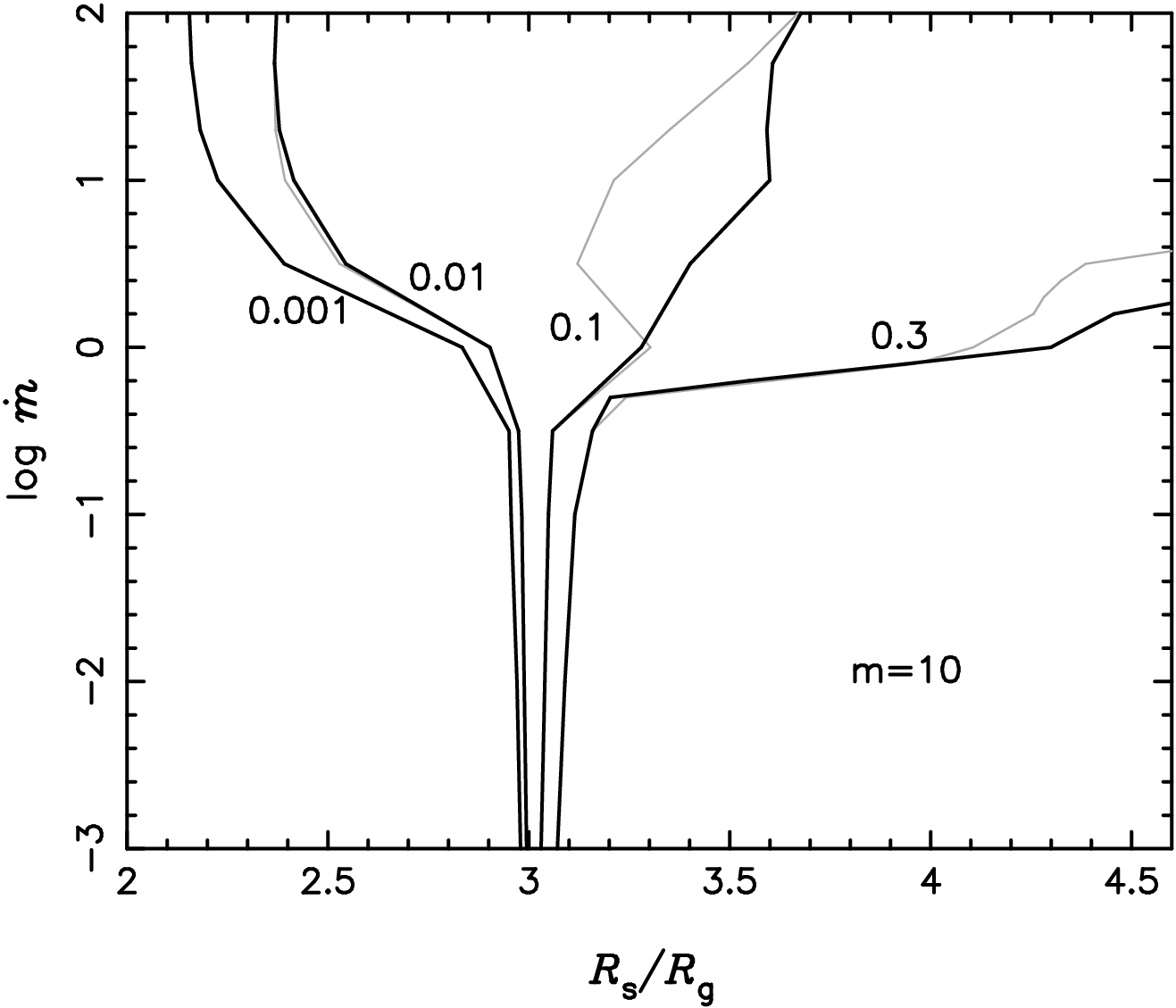}}
\figcaption{Location of the transonic point ($R_\mathrm{s}$) in function of the accretion rate
($\dotm$) and the viscosity parameter ($\alpha$; labeled for each line).} 
\label{sonicpoint}
\vglue 0.2cm

\figurenum{4}
\begin{figure*}[tb]
\centerline{\includegraphics[angle=-90.0,width=17.5cm]{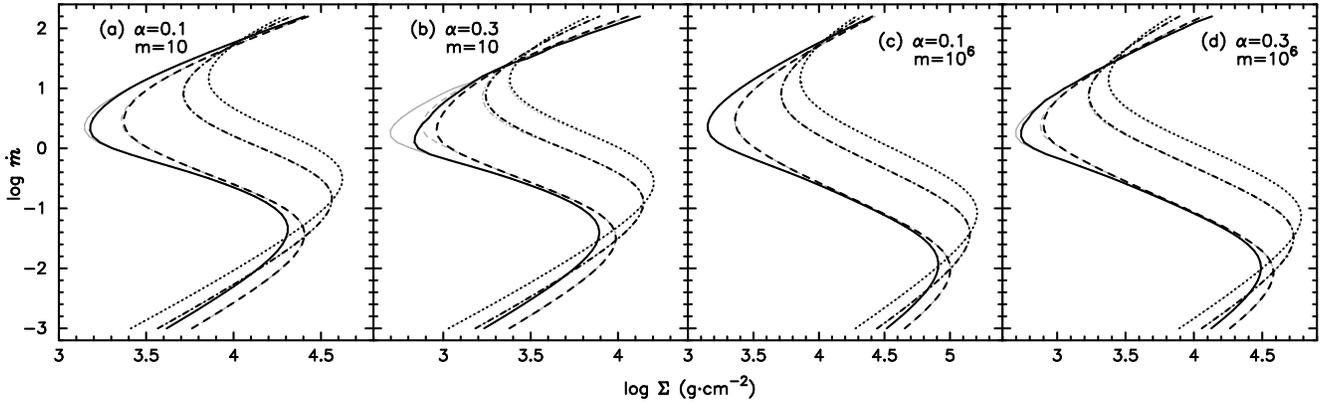}}
\figcaption{S-shaped curves for accretion flows of stellar black holes and SMBHs. The solid,
dashed, dot-dashed and dotted lines are at the radii $5$, $10$, $50$ and $100$ $R_{\rm g}$,
respectively.}
\label{sshape}
\end{figure*}
\vglue 0.1cm

%\noindent 
The detailed transonic situations are analysed here, and they are compared with the cases of
slim disks without the correct treatment of transition regions. Fig. 2 presents the exact
distributions of the angular momentum ($l$), totoal pressure ($P$) and radial velocity
($\upsilon_\mathrm{R}$) in the vicinity of transonic position, with the viscosity parameter
$\alpha=0.1$ and $m=10$. The cases of $\dotm=1$ and $10$ are both exhibited. From this figure,
it can be seen that the curves of $\dotm=1$ of our model are almost the same as those of the
classical slim disk model. Comparing with the classical slim disk, in the slim disk with transition
regions, the angular momentums shift upward, total pressures decrease
and the ratios of radial velocity to sound speed increase at the innermost radius, due to
the effects of optically thin region. In the inner disk region where
the effective optical depth is smaller than unity, the diffusion approximation for blackbody
radiation is invalid, and the radiative cooling becomes inefficient. Thus the temperature is
higher than that of the classical slim disk, where local blackbody radiation is assumed in the
entire disk.
At the same time, the lower total pressure resulting from the decreased radiation pressure due
to $\tau_\mathrm{eff}<1$, leads to the smaller viscous tensor ($\tau_{r\varphi}\propto{P}$).
Therefore, the curve of angular momentum is flatter (see equation 3) and the flow
becomes more dense. The accretion flow reaches the sonic point earlier than the classical slim disk,
since the sound speed $c_\mathrm{s}=(P/\rho)^{0.5}$ is much smaller and the ratio
$\upsilon_\mathrm{R}/c_\mathrm{s}$ becomes larger. This explains why the transonic positions shift
outwards, compared with those in Abramowicz et al. (1988).

As well known, the accretion flow has four types of transonic cases, according to which kind of
pressure (gas or radiation) is dominant at the sonic point and the magnitude of the
viscosity parameter. For low $\alpha$, it is total pressure that pushes the flow pass through
the sonic point. While for large $\alpha$, the action of viscosity drives the flow across the
transonic point. Fig. 3 shows the dependency of the transonic radius on the accretion
rate and the viscosity parameter for a black hole of 10 $M_\odot$. It is apparent that the
transition to an optically thin region moves the transonic point out, when the accretion rate is
in the range $1<\dot{m}<100$ and the viscosity parameter is $\alpha\ge{0.01}$. In other parameter
space comprising $\dotm$ and $\alpha$, the transonic location in our result is almost the same as
that of the classical slim disk.

The transonic flows with transition regions of high viscosity can be compared with the flows
of low viscosity. First, they are both radiation pressure dominated for super-critical
accretion rates. Second, they are different in some properties. One of them is manifested
in the optical depth. The accretion flow is almost entirely optically thick for 
a small viscosity parameter, e.g. $\alpha=0.0001$. While the transition region appears for a large
$\alpha$ (if $\alpha$ is large enough, the inner flow can be purely optically thin and gas-pressure
dominated). Another distinct feature lies in the position of the sonic point. For low $\alpha$, 
the flow cannot acquire enough velocity to transonic until it is near the marginal bound orbit
(2 $R_\mathrm{g}$), since the sound speed is relatively high and the viscosity-driven
process is not efficient. On the contrary, the sound speed of the flow with large $\alpha$ 
is small, the sonic point locates far outside of the last stable orbit (3 $R_\mathrm{g}$). 
Therefore, we conclude that the transonic situation of the accretion flow with transition region is,
to some extent, between the case of low viscosity and that of high viscosity in the classical slim disk. 

The S-shaped curves for different parameters are shown in Fig. 4. It is found that they are
strongly affected by the transition region when the viscosity is high. 
For the case of stellar-mass black holes, the upper ADAF and the middle unstable branches
change much. The upper branch due to advection is modified, and the unstable branch is compressed.
The appearance of the transition to an optically thin region lowers the radiation efficiency,
hence the advection becomes more important. In the accretion flows of SMBHs, the S-shaped curves
are less altered, as shown in panels (c) and (d) in Fig. 4. It is easy to understand the curves
are not seriously influenced in the outer region, since there the optical depth is large and
it tends to be the classical slim disk. The distorted S-shaped curves may have observational features
in stellar black hole accretions. The transition timescale from the upper to the lower branches,
$t_\downarrow$, will be shorter than that of the classical slim disk. For example with
the same parameters,
the difference at $R=5R_\mathrm{g}$ has a factor of $2-4$. But the transition timescale from
the lower branch to the upper, $t_\uparrow$, is unchanged and greater than $t_\downarrow$.
This would be of interest in the explanation of X-ray transient objects. The variation
of state-transition timescales in stellar black holes is also one of the crucial characteristics
of the transition to optically thin regions in slim disks.

\figurenum{5}
\begin{figure*}[htb]
\centerline{\includegraphics[angle=-90.0,width=17.0cm]{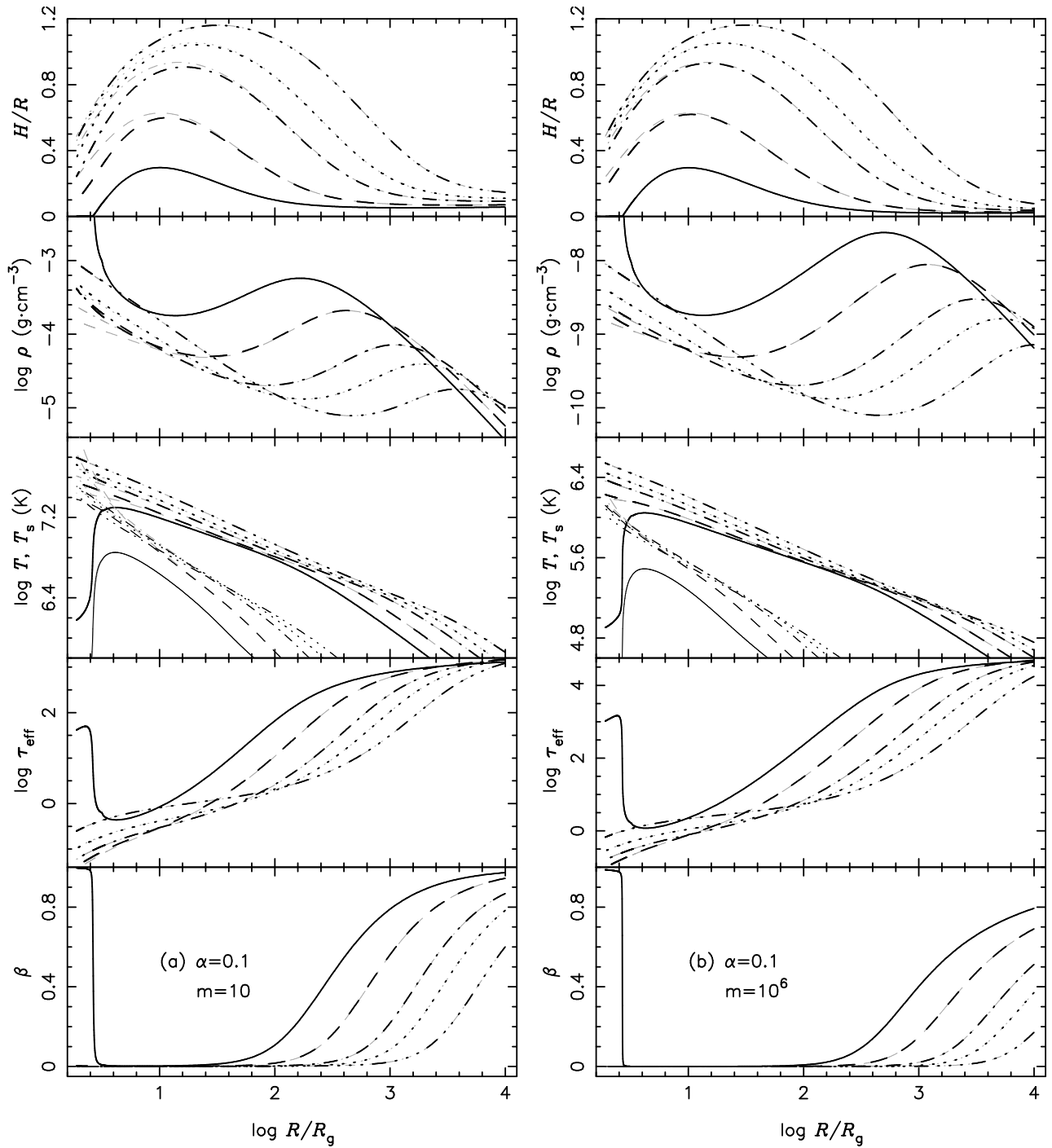}}
\figcaption{$\dot{m}$-dependency of the aspect ratio ($H/R$), density ($\rho$), midplane temperature
($T$) and surface temperature ($T_\mathrm{s}$), effective optical depth ($\tau_\mathrm{eff}$), ratio
of gas pressure to total pressure ($\beta$) for black holes of (a) $m=10$ and (b) $m=10^6$ with
$\alpha=0.1$. The solid, dashed, dot-dashed, dotted and dot-dot-dot-dashed lines are for
$\dot{m}=1$, $3$, $10$, $20$ and $50$, respectively. The surface temperatures $T_\mathrm{s}$
are shifted down by 0.5 logarithmic units for clarity.}
\label{str-mdot}
\end{figure*}

\subsection{Global disk structure}
Fig. 5 provides the physical quantities such as the half-thickness, density,
midplane temperature (including surface temperature), effective optical depth and the ratio of gas
pressure to total pressure for slim disks of both stellar-mass black holes and SMBHs.
The results of the classical slim disks are also shown in the figure with grey lines.
The transition region influences greatly the structure in inner disk region, especially for
the accretion flow around a stellar-mass black hole.

First is the half-thickness. For all the super-critical accretion rates $1\le\dotm\le50$,
$H\lesssim1$, indicating the flow is geometrically slim. The thickness increases with the
accretion rate, which is consistent with the result of the classical slim disk.
At the outer boundary radius,
the disk is always geometrically thin, and the solutions can be approximated by Shakura-Sunyaev
solutions. Because of $H=B_2c_\mathrm{s}/\Omega_\mathrm{K}$, the lower sound speed results in
the reduction of half-thickness, in comparison with the classical slim disk model. The reduced height
favors the trapped photons to escape from the disk. This shape of the accretion flow is somewhat
analogical with the funnels in thick disks. 

With the emergence of the transition region, the temperature and density of the disk are both
larger than those in the classical slim disks. There are still considerable radiation fluxes
in the innermost region after the flow passes through the sonic point, which is pointed out
by Watarai et al. (2000). The effective optical depth has also been shown in Fig. 5.
It should be noted that there is a large zone of $\tau_\mathrm{eff}\le1$ when $\dotm\ge3$,
e.g. more than 30 $R_\mathrm{g}$ (15 $R_\mathrm{g}$) for the disk of a black hole of $10$ ($10^6$)
solar masses. The lowest effective optical depth reaches 0.03. More important is that,
in the accretion flows of both stellar-mass and supermassive black holes, there are wider regions
of at least 100 $R_\mathrm{g}$ showing $\tau_\mathrm{eff}\sim\/1$ for a moderate super-critical
accretion rate $10\le\dotm\le50$. In such large transition regions, blackbody radiation is not
adequate to describe the radiative transfer. Accordingly, a new radiation formula valid for any
optical depth, as adopted in the present work, is necessary.

When the accretion rate tends to very large, for example $\dotm>10$, in the inner disk region
the advected energy dominates over the surface cooling, i.e. $Q_{\rm adv}\gg Q_{\rm rad}$,
the flow will have self-similar behaviors (Wang \& Zhou 1999). We find that the global solution
indeed shows the self-similar signatures: the radial velocity $\upsilon_{\rm R}\propto r^{-1/2}$,
total pressure $P\propto r^{-5/2}$, the density $\rho\propto r^{-3/2}$, the temperature
$T\propto r^{-5/8}$ and $\tau_{\rm eff}\propto r^{3/16}$ weakly depends on the radius, except
in the vicinities of the inner disk edge. Watarai \& Mineshige (2003) have also shown
these properties.

\figurenum{6}
\centerline{\includegraphics[angle=-90.0,width=8.5cm]{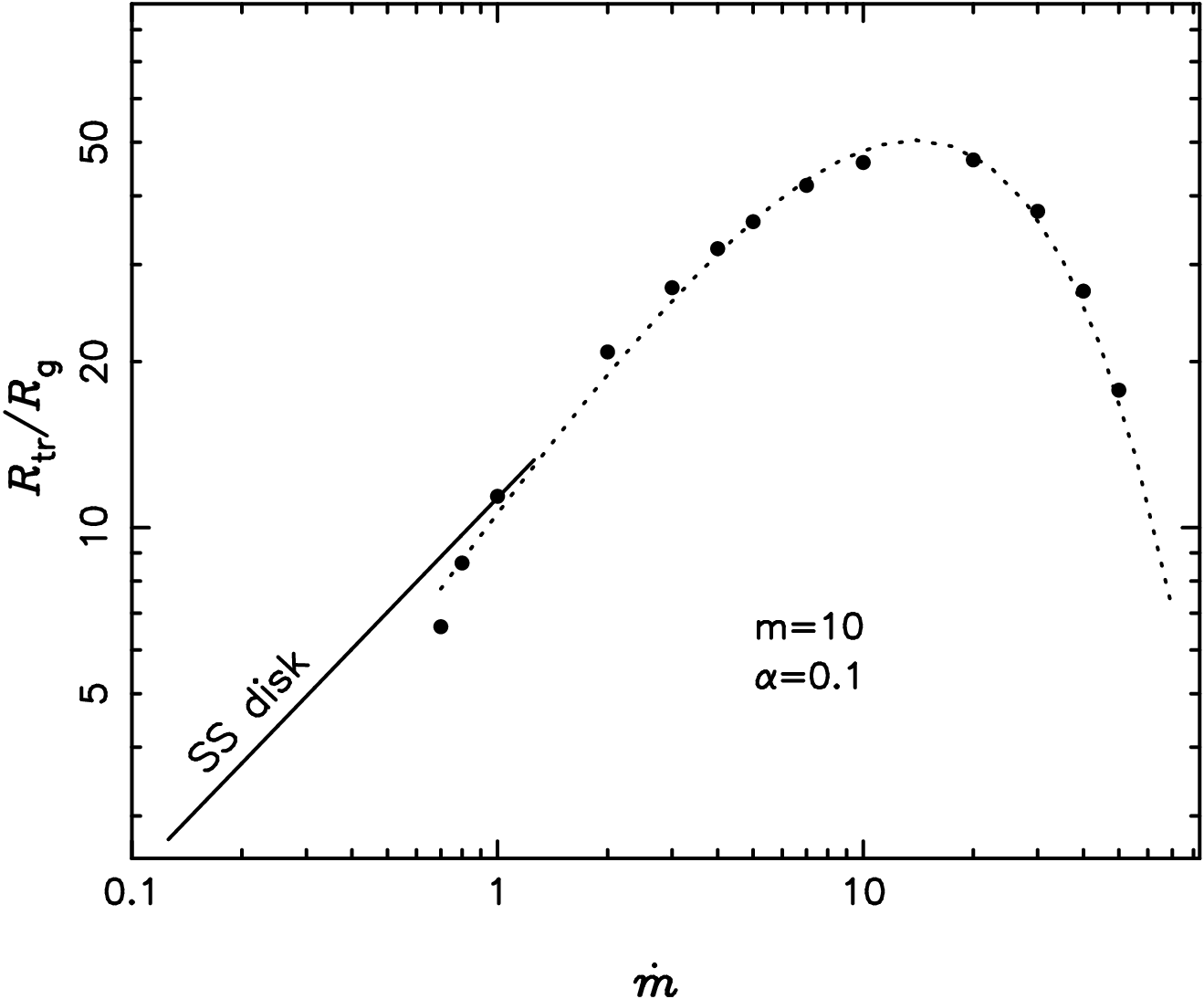}}
\figcaption{The dependence of the transition radius on the accretion rate for a fixed viscosity
$\alpha=0.1$ in stellar black holes. The black points are the results from our slim disk model. There
is a maximum radius of the transition region. The solid line is equation (1) for Shakura \& Sunyaev
(1973) disk. The dotted is the fitting line by the least square method.}
\label{dotm-tr}
\vglue 0.2cm

Fig. 6 shows the dependence of the transition region on the accretion rate.
We find the relation can be well fitted by an analytical formula
\begin{equation}
R_\mathrm{tr}/R_\mathrm{g}=95.38\alpha^{0.91}\dot{m}^{0.96}\mathrm{exp}\left(-0.1\dot{m}^{0.9}\right).
\end{equation}
There is a maximum transition radius $R_{\rm tr,max}/R_{\rm g}\approx 50$
at $\dot{m}\sim 15$. This is clearly different from that in the standard disk model, i.e.
$R_{\rm tr}\propto \dot{m}^{64/93}$ (see equation 1). Our calculations show that there is no
transition radius when $\dotm<0.5$, and it drops quickly if $\dot{m}>30$.
The maximum radius of the transition region can be understood from the S-shaped curves in Fig.
4. Near the turning point between the upper and the middle branch, the surface density
is the lowest and at the same time the very high temperature makes the absorption opacity be
small. The father from this area, the larger effective optical depth due to the increase of either
the surface density or the absorption opacity, or of both.
This is an interesting result since the intermediate super-critical accretion has the largest
transition radius, where the gravitational energy is mostly released.

We briefly discuss the viscosity effect on the disk structure. Fig. 7 is the
$\alpha$-dependency of the global disk structure. In the slim disks, $\alpha$ has only a little
effect on the disk half-thickness. In our model large $\alpha$ such as $0.3$, reduces much
height, increases the density and temperature in the inner region, which differs with
that discussed by Kawaguchi (2003). At the outer radius, the situation is similar as the result in
Kawaguchi (2003): the larger viscosity, the lower density, midplane temperature and also the
optical depth, but the higher surface temperature. This is because of more efficient angular momentum
transfer outwards and more trapped advection energy in the flow.

\figurenum{7}
\centerline{\includegraphics[angle=-90.0,width=8.5cm]{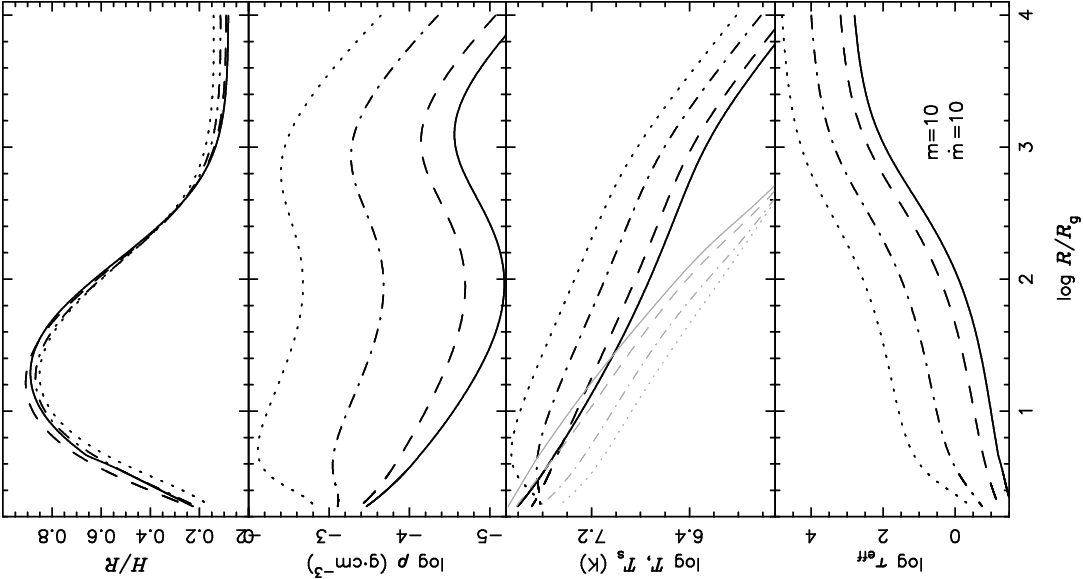}}
\figcaption{$\alpha$-dependency of the physical quantities for $m=10$ and $\dotm=10$. Solid, dashed,
dot-dashed and dotted lines are for $\alpha=0.3$, $0.1$, $0.01$ and $0.001$, respectively. The surface
temperatures ($T_\mathrm{s}$; grey lines) are shifted downward by 0.2 logarithmic units.}
\label{str-alpha}
\vglue 0.2cm

Fig. 8 is the dependence of the transition radius on the viscosity $\alpha$. It can 
be found that the results from our calculations are not in agreement with the Shakura-Sunyaev disk,
even at a small viscosity ($\alpha<0.01$). This is because that the Shakura-Sunyaev disk breaks
down when $\dotm>0.2$. The dependence of the transition radius on $\alpha$ becomes very strong,
roughly $R_{\rm tr}\propto \alpha^{0.91}$ for $\alpha>0.01$ and $R_{\rm tr}\propto \alpha^{0.22}$
for $\alpha<0.01$, while $R_{\rm tr}\propto \alpha^{34/93}$ in the Shakura-Sunyaev disk.
The high viscosity drives the accreting gas to fall into the black hole more rapidly, leading
to a low density region. This region then becomes absorption weak, enlarging the transition
region. Since the radial motion and the advection effect is neglected in the standard disk,
its prediction does not hold when either the accretion rate or the viscosity is large. 

In the end, we should pay attention that the properties of the effective optically thin region
discussed here are completely different from optically-thin ADAFs. Firstly, the flow is only effective
optically thin. The volumn density and the surface density are still larger than those of
optically-thin ADAFs. The Thomson scattering is very important and the Comptonization is significant.
That the optical depth becomes effectively thin is because of the transparancy to photon absorption at
a high temperature. In optically-thin ADAFs, the plasma density is very low and the Thomson
scattering optical depth is rather small. Secondly, the temperature in the present slim disk model
is not so high as that of optically-thin ADAF ($10^{8-9}$ K). Thirdly, the pressure is still
radiation-pressure dominated. While for the ADAF, it has merely gas pressure, the radiation pressure
can be neglected. However, we believe that slim disks (optically-thick ADAFs) are possible to transit
to optically-thin ADAFs in inner region, once the viscosity is large enough. We leave this topic
in the later paper.

\figurenum{8}
\centerline{\includegraphics[angle=-90.0,width=8.5cm]{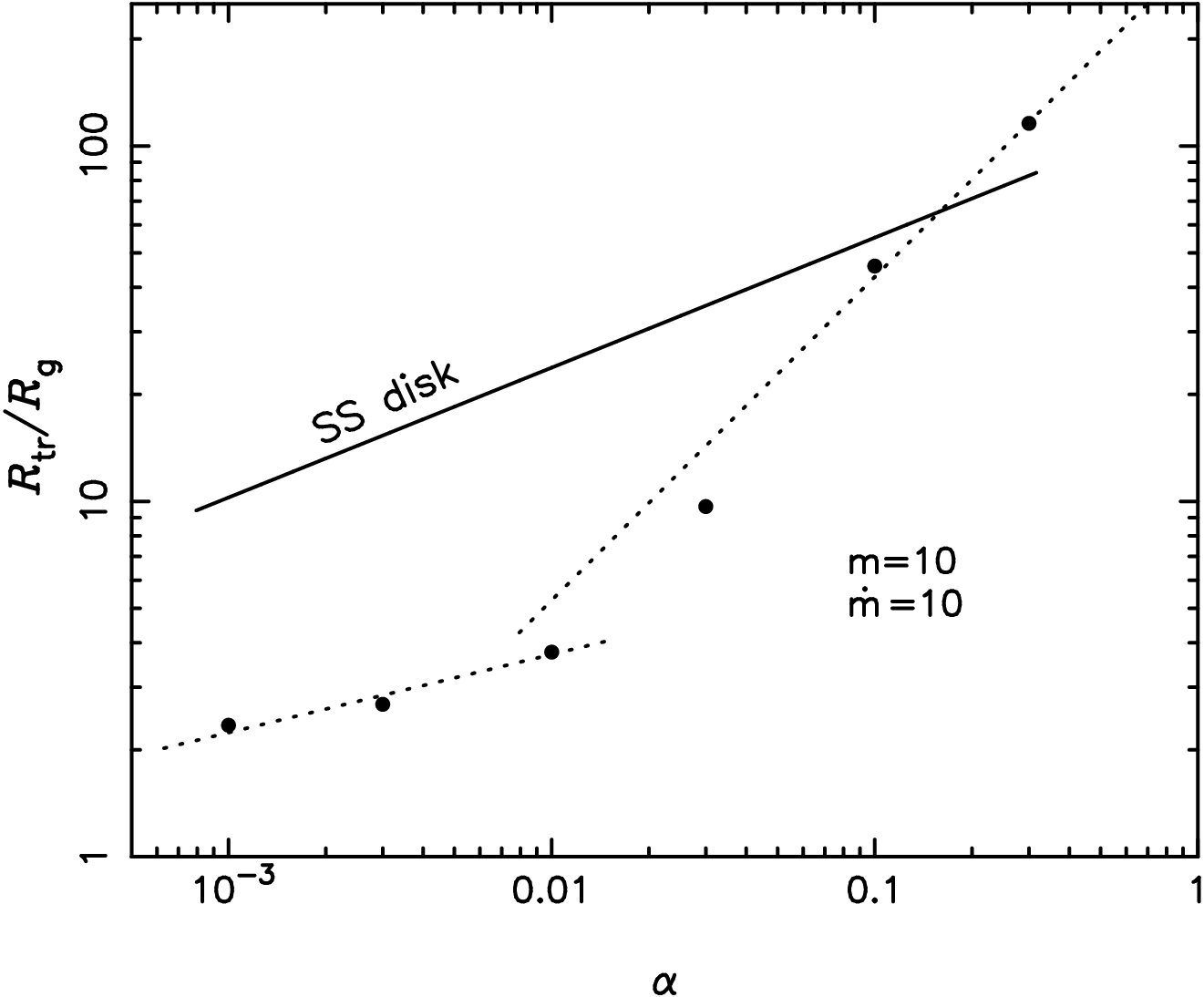}}
\figcaption{The dependence of the transition radius on the viscosity for a fixed accretion rate
$\dotm=10$ in stellar black holes. The black points are the results from our slim disk model.
The solid line represents the equation (1) from Shakura \& Sunyaev (1973) model. The dotted line is
the fitting line.}
\label{alpha-tr}
\vglue 0.2cm

\figurenum{9}
\begin{figure*}[tb]
\centerline{\includegraphics[angle=-90.0,width=17.0cm]{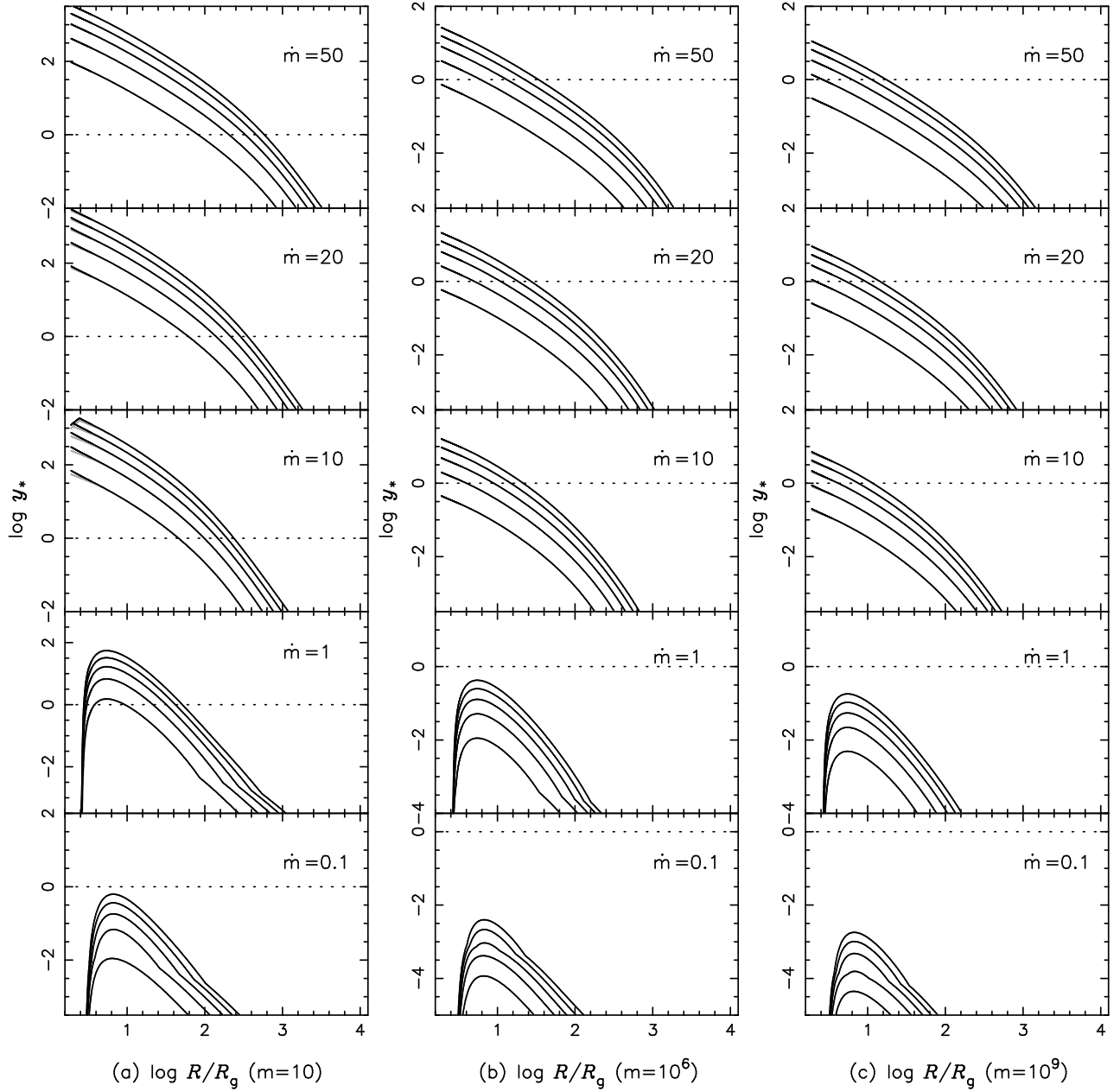}}
\figcaption{Effective Compton parameter $y_*$ in slim disks with transition regions. In each panel,
the solid lines correspond to $x=h\nu/k_BT=0.2$, $0.4$, $0.6$, $0.8$ and $1.0$, respectively,
from bottom to top.}
\label{compton-y}
\end{figure*}
\vglue 0.2cm

\subsection{Emergent spectra}

\figurenum{10}
\begin{figure*}[tb]
\centerline{\includegraphics[angle=-90.0,width=17.5cm]{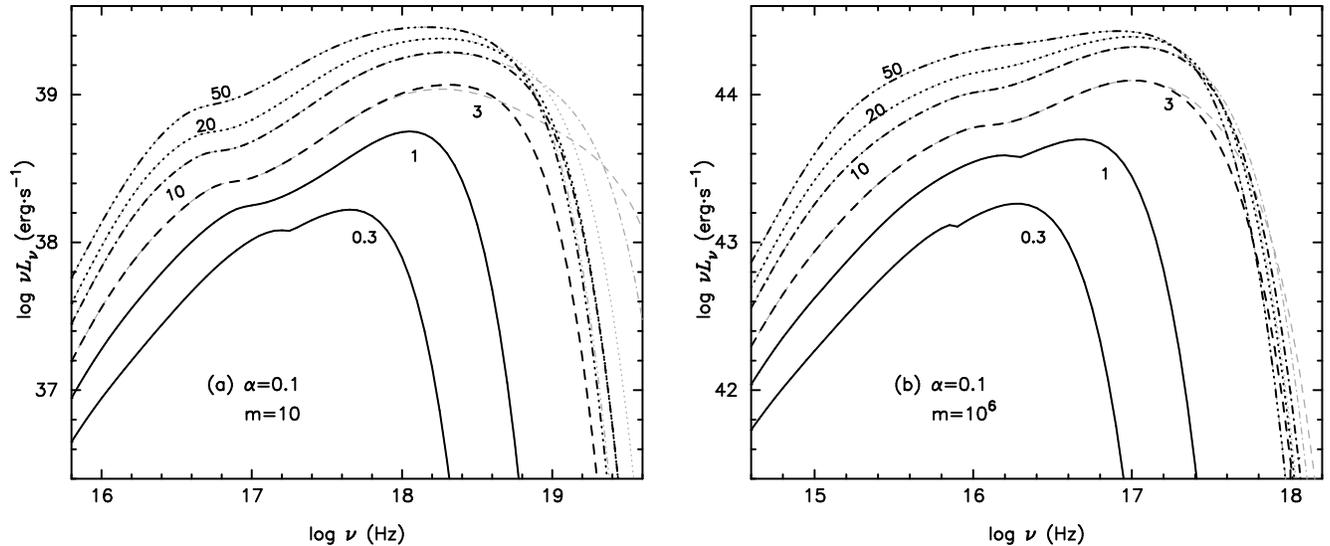}}
\figcaption{The emergent spectra from slim disks with transition regions for (a) $m=10$ and (b)
$m=10^6$. From bottom to top, the accretion rates are $0.3$, $1$, $3$, $10$, $20$ and $50$,
respectively. The grey lines are the results for the classical slim disks, of which the calculated
surface temperature is higher than the midplane value in the optically thin region. Thus the
radiation fluxes at high energies are overestimated.}
\label{spectra-mdot}
\end{figure*}
\vglue 0.2cm

\figurenum{11}
\begin{figure*}[tb]
\centerline{\includegraphics[angle=-90.0,width=17.5cm]{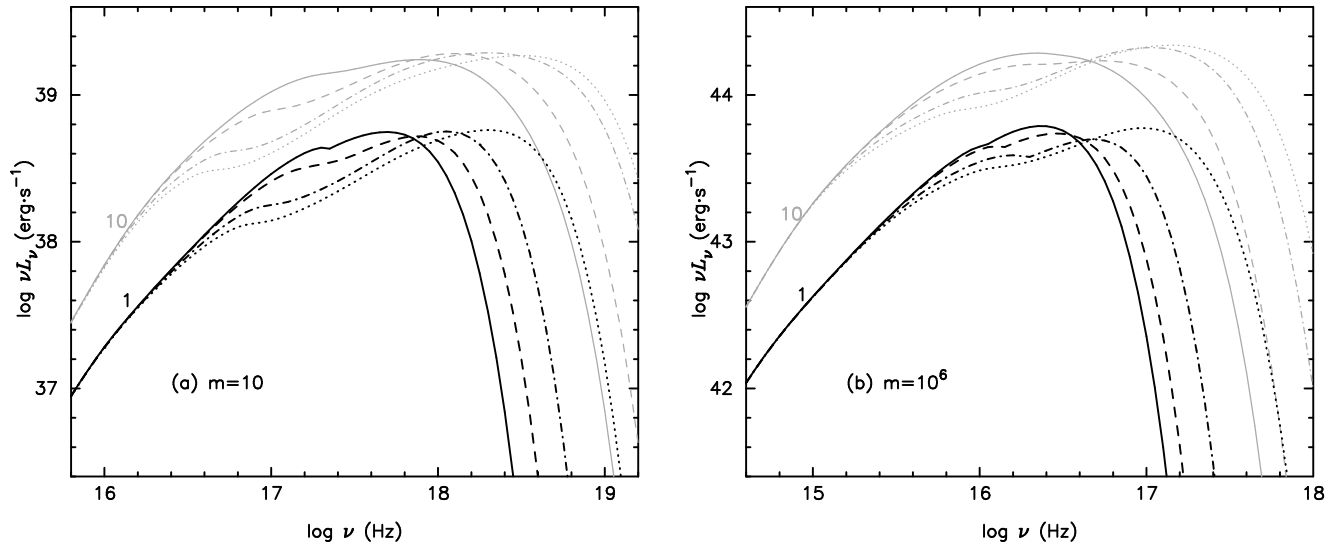}}
\figcaption{$\alpha$-dependency of the emergent spectra from slim disks with transition regions. 
The solid, dashed, dot-dashed and dotted lines are for $\alpha=0.001$, $0.01$, $0.1$ and $0.3$, 
respectively. The accretion rates are labeled in each panel.}
\label{spectra-alpha}
\end{figure*}
\vglue 0.2cm

We have shown that the global structures of the slim disks with transition regions are quite
different from those slim disks in Abramowicz et al. (1988). With the presence of transition
region, only the theoretically predicted spectrum based on the global structure is of great
significance to the observations.

The emergent spectra of slim disks have been extensively studied. Szuszkiewicz et al. (1996)
were the first to calculate the spectrum with modified blackbody radation, and explained
the soft-X excesses of quasars. Wang et al. (1999) calculated the vertical structures
and emergent spectra from slim disks. Two prominent features appear: the saturated total
luminosity and the constant cutoff energy. Mineshige et al. (2000) and Watarai et al. (2001)
used purely blackbody radiation approximation, though it is not adequate in inner disk region.
Wang \& Netzer (2003) and Kawaguchi (2003) employed the method of Czerny \& Elvis (1987) to
consider Thompson electron scatterings and Comptonization in accretion disks.
However, since they did not treat the phenomena of optically thin, their results are
only available for low viscosity parameter. As for the case of high viscosity parameter,
the local radiation in spectral calculations tends to thermal bremsstrahlung when the optical
depth is small, while local blackbody radiation that holds only for optically thick medium
is assumed to determine the disk structure. This inconsistent treatment between structure and
spectral calculations leads to the problem of temperature invertion (Czerny \& Elvis 1987),
namely, the calculated surface temperature is even higher than the midplane temperature in the
accretion flow (see the grey lines in the middle panels of Fig. 5; In the classical slim disk,
$T_\mathrm{s}$ is much larger than $T$ at several $R_\mathrm{g}$). The reason is that
the blackbody approximation predicts too much energy loss than the realistic radiation in
optically-thin flows. Thus, the radiated flux is heavily overestimated at high energies.

We have solved the problem in this paper. By adopting the bridged formula which can describe the
radiation energy for both optically-thick and -thin cases, the calculated surface temperature is
not larger than the midplane temperature, as shown in Figs. 5 and 7. Except that the Thompson
electron scattering modifies the radiation to deviate blackbody, the inelastic scattering also
exchange the energy between photons and electrons. The Comptonization effect has been stressed
in Wang \& Netzer (2003) and Kawaguchi (2003). We can give a more detailed estimation of the
importance of this process. The effective Compton parameter 
\begin{equation}
y_*=\Xi_e\mathrm{Max}(\tau_\mathrm{es}^\prime,\tau_\mathrm{es}^{\prime2}),
\end{equation}
where $\tau_\mathrm{es}^\prime=\tau_\mathrm{es}/\mathrm{Max}(\tau_\mathrm{eff},1)$, the absorption
opacity is taken as $\kappa_{\mathrm{abs},\nu}=1.4\times10^{25}\rho\/T^{-3.5}x^{-3}(1-\mathrm{e}^{-x})$
($x=h\nu/k_BT$), and 30 times value for SMBHs. Fig. 9 shows the parameter $y_*$ for different
photon energies, accretion rates and black hole masses. When $y_*>1$, Comptonization can not be
neglected and the emergent spectrum will be strongly modified. Fig. 9 clearly shows how important
the Comptonization process is for disks with different parameters. We can draw a conclusion that
the Comptonization effect is very important in the inner region of super-critical accretion flows.
Saturated Comptonization process is inevitable for less massive black hole disk ($10^{6-7}M_{\odot}$).
This confirms the conclusion of Wang \& Netzer (2003). The Comptonization is less important for
a $10^9M_{\odot}$ black hole disk unless the disk has highly super-critical accretion rate,
which agrees with Laor \& Netzer (1989). The higher accretion rate, the more important
Comptonization process. We adopt the scheme described in \S2.2 to calculate the spectra from
slim disks, taking into account the Comptonization effect by a simple but practical method.

Figs. 10 and 11 show the details of the emergent spectra from the slim disks with transition
regions. We would like to stress that the present calculations have two advantages: 1) we include
the transition region; 2) the model covers wide ranges of the accretion rate and the viscosity.
Thus the present model involves all the known disk models (except for the optically thin ADIOs
and CDAFs). Fig. 10 clearly shows the spectra from the accretion disks of which the central black
hole masses are 10 and $10^6$ $M_{\odot}$. We start the accretion rate from sub-critical
($\dot{m}=0.3$) to super-critical ($\dot{m}=50$). We find the spectrum shifts toward high energy
with the increasing accretion rate. This is due to the increases of temperature and of the importance
of Comptonization with the accretion rate. The spectra show very strong humps at high energies
for both stellar and supermassive black holes. The spectral flux is proportional to the accretion
rate for the sub-critical accretion flow. The distinct spectral characteristic of super-critical
disks is its broadness and flatness. The cutoff energies of the spectra almost keep constant
when the accretion rate $\dot{m}$ is above 10, and the total radiated luminosity from the disk tends
to be saturated. The {\em constant} cutoff energy was first found in Wang et al. (1999), which is
caused by {\em advection}. The feature of the saturated luminosity has also been found in
Abramowicz et al. (1988), which is caused by the advection effect: most of the dissipated energy
is swallowed by the central black hole. The maximum cutoff energy is about 100 keV (4 keV) for
a black hole of 10 ($10^6$) solar masses. The saturated luminosity is about several Eddington
limit ($\sim10^{39}$ and $10^{44}$ erg/s for 10 and $10^6$ $M_\odot$ black holes, respectively).
This extreme slim disk with very high accretion rate can be explained by the behaviors of
self-similar solution (Wang \& Zhou 1999; Wang \& Netzer 2003). The features of the broad and
flat spectrum, constant cutoff energy and saturated luminosity are very useful in explaining
the X-ray spectra of microquasars, ULXs and narrow line Seyfert 1 galaxies.

For a comparison with the spectra from the classical slim disks, we draw them as grey lines in
Fig. 10. We use the same computation scheme for the spectra from the classical slim disks.
We find that the emergent spectra are quite different from those based on the present model with
transition region, when the accretion rate is above the critical value. The cutoff energy in the
classical slim disk is unrealistically higher than the present model. Thus the present model
provides an improved version for the slim disk.

We investigate the effect of the viscosity on the spectrum. For all the accretion rates,
larger viscosity parameter causes the spectrum to extend to higher frequency. This is the
contribution of the more significant elastic electron scatterings and Comptonization. 
From Czerny \& Elvis (1987), the critical temperature
is $T_\mathrm{es}\propto\dotm^{-4/11}m^{1/11}\alpha^{-3/11}$, above which the modification due to
electron scattering is important. If $\alpha$ increases, the transition region will enlarge and the
surface temperature will increase. In such a flow the Comptonization effect becomes stronger.
Thus the emergent spectrum is broader and flatter with the higher viscosity parameter. 
Usually the cutoff energy strongly relies on both the accretion rate and the viscosity parameter.
But in a saturated accretion system, it is only affected by $\alpha$. This may be utilized to
determine the unknown viscosity parameter through X-ray observations of ultra-luminous objects.

Finally, with the detailed calculations, we verify the importance of photon trapping in the 
super-critical flows.  If the diffusion timescale
\begin{equation}
t_\mathrm{diff}=\frac{H(\tau_\mathrm{es}+\tau_\mathrm{abs})}{c},
\end{equation}
is longer than the accretion timescale
\begin{equation}
t_\mathrm{acc}=\frac{R}{\upsilon_\mathrm{R}}
\end{equation}
in a flow, the photons can NOT escape from the disk at the same radius. This causes the so-called
photon trapping. Such a case has been recognized in early time (Katz 1977; Begelman 1978), but the
quantitative features have been discussed by Wang \& Zhou (1999); Ohsuga et al. (2002, 2003) and Shimura \& Manmoto (2003).  

Let us compare the two timescales. Fig. 12 shows $t_{\rm acc}\gg t_{\rm dyn}$ in the
entire $r\le 10^4$ region for any $\dot{m}\le 50$. This means that the viscosity works for the
disk, transporting the angular momentum and dissipating the gravitational energy. When $\dotm>1$,
in the inner region $t_\mathrm{diff}>t_\mathrm{acc}$, the photons are hardly radiated out
so that trapped in the accretion flow. In this trapped region, energy transport in the form of
advection is the main cooling mechanism. In this sense, the slim disk is often called optically 
thick ADAF. After comparisons, we conclude that the photon trapping effect is still significant,
but it is not so severe as that in the classical slim disk. This is mainly because of the greatly
reduced disk height. The present model does not treat the photon trapping carefully, but see
Ohsuga et al. (2002, 2003), who treated the photon trapping with the given disk structure. 
Wang \& Zhou (1999) have shown that the radiated luminosity is of 
$4\times 10^{38}(M_{\rm BH}/10M_\odot)$ erg/s for an extreme super-critical disk,
which is as small as one third of the Eddington luminosity. In fact, in physics the photon trapping
effect makes it difficult to estimate the real accretion rate from the saturated observed spectrum,
when the black hole has a super-critical accretion rate. One of the goals in this
paper is to investigate the physics of the transition region. For a moderate super-critical
rate $\dot{m}<10$, the photon trapping effect is not so strong, our present model does work for
our applications in next section. The model coupling the photon trapping effects is just in preparation. 
Considering the photon trapping effect, we believe that the super-critical accretors in microquasars
and NLS1s are more than the number we know at the present. Kawaguchi (2004) realized the role of
the self-gravitation in the optical spectrum of NLS1s. We did not consider the self-graviation in the
present work. The general relativistic effects are not included either. Future work will
solve the equations of the general relativistic disk with transition region, as well as calculate
the spectrum from the disk.

\figurenum{12}
\centerline{\includegraphics[angle=-90.0,width=8.5cm]{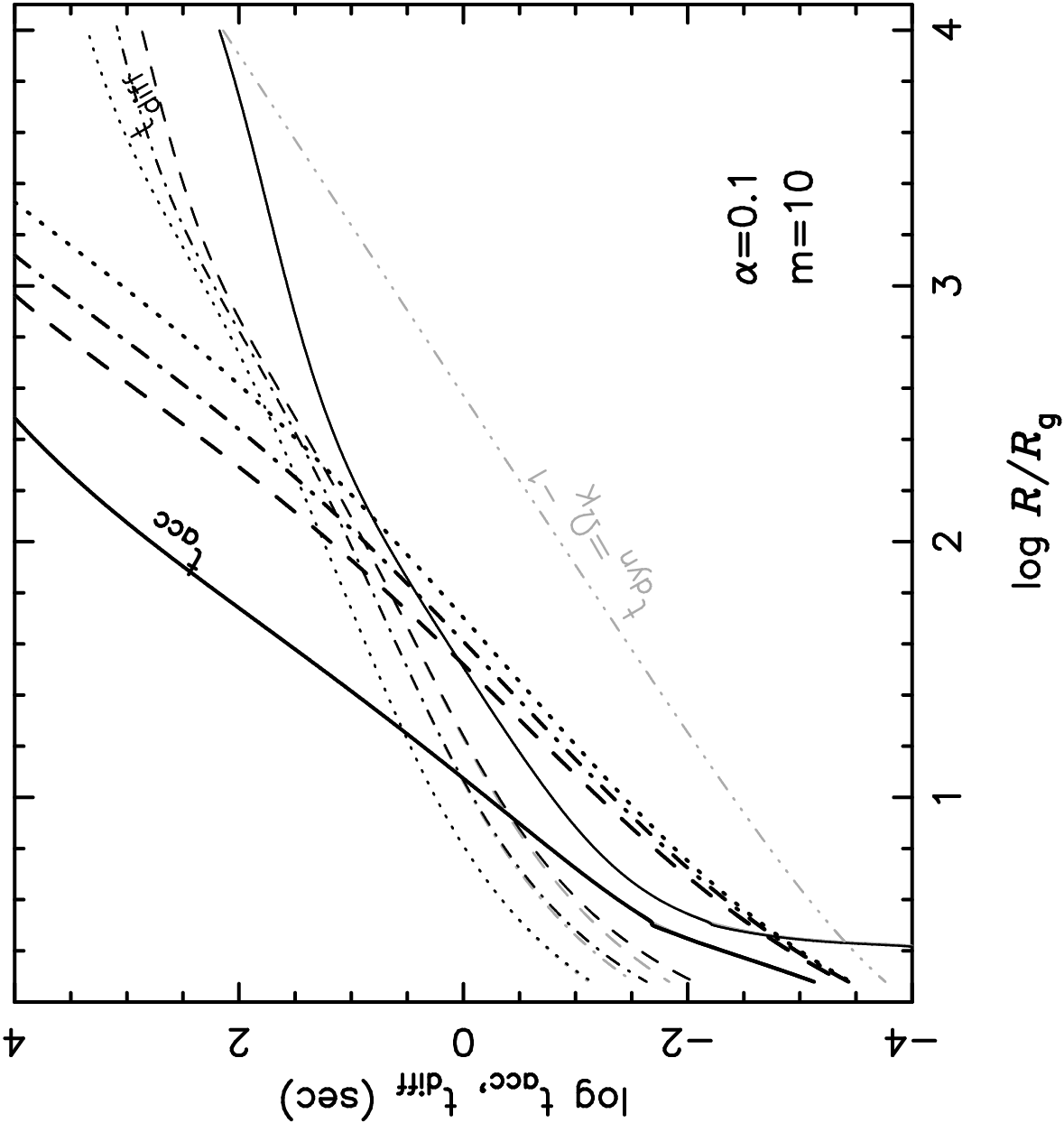}}
\figcaption{Accretion timescale ($t_\mathrm{acc}$; thick lines), diffusion timescale
($t_\mathrm{diff}$; thin lines) and dynamic timescale ($t_\mathrm{dyn}$). The solid, dashed,
dot-dashed and dotted lines are for $\dotm=1$, $10$, $20$ and $50$, respectively.
The grey lines are the results of the classical slim disks.}
\label{timescale}
\vglue 0.2cm

\section{Applications}
The most prominent characters of the spectra from slim disks are the strong humps and
the constant cutoff frequency. When $L/L_{\rm Edd}\ge 0.3$, the standard accretion disk model
breaks down (Laor \& Netzer 1989) and the slim disk is available. Our present model covers a
rather wide range of accretion rates. The distinguished soft X-ray humps in some objects are
explained by the self-consistent emergent spectra based on the corresponding global disk
structures. In this section, we apply our model of slim disks to the microquasar GRS 1915+105
and two NLS1s: RE J1034+396 and Akn 564.

\subsection{Microquasars: GRS 1915+105}
Some of galactic black hole candidates, especially microquasars and ULXs, are regarded to be
super-critical accretors (Watarai et al. 2000, 2001). The broad band X-ray spectra of
microquasars and ULXs have been observed extensively with {\rxte} and {\asca}
(Makishima et al. 2000; Mizuno et al. 2001; Ueda et al. 2002). The observed spectra are fitted
well with multicolor blackbody spectra in the framework of standard optically-thick accretion disk.
However, the multicolor disk (MCD) spectrum has some shortcomings. The spectral hardening factor
due to Comptonization in the disk atmosphere is assumed to be a constant (usually 1.7;
Shimura \& Manmoto 2003) over the entire disk. Our calculations show Comptonization is only crucial
in inner region. Additionally, MCD is based on the standard disk theory, the viscous dissipation is
thought to be radiated efficiently in the local. This endows an upper luminosity limit
($L<0.3L_\mathrm{Edd}$; Laor \& Netzer 1989) to its use scope. Obviously, the luminosities of
ULXs and other bright objects are beyond this limit. Therefore, the fitting parameters of MCD
model have difficulties to connect with the physics. First, the X-ray estimated black hole mass
with MCD is smaller than the precise dynamic measurement by optical observations. Second,
the fitted color temperature is too hot to be produced by the standard disk, i.e. the problem
of too hot disk (e.g. King \& Puchnarewicz 2002; Ebisawa et al. 2003). Lastly, the MCD model
cannot account for spectral variations. The five periods of spectral fitting of IC 342 source
1 (one ULX) show that the black hole mass varies by a factor of 2 (Ebisawa et al. 2003).
This is unreasonable.

Ebisawa et al. (2003) found the spectra of Galactic superluminal jet sources can be explained
with standard Kerr disk model, while ULXs need slim disk model. Sobolewska \& \.{Z}ycki (2003)
stated that the relativistic effects of Kerr model can only partially account for the apparently
complex soft X-ray spectra of GRS 1915+105. In this subsection, we show that the emergent spectrum
from our slim disk model, considering the effects of deviated blackbody radiation, Comptonization 
and the transition to optically thin, is quite adequate to describe the soft spectral
component of GRS 1915+105 when it is in high state. 

The measurements of the black hole mass and the distance in GRS 1915+105 are quite reliable, so
it is able to apply the present model to this object. The black hole mass $M_{\rm BH}=(14\pm4)M_\odot$
is deduced from the direct measurement of the orbital period and mass function of GRS 1915+105
(Greiner et al. 2001). The inclination is $i=70^\circ$, the distance ${\cal D}=12.5$ kpc
(Mirabel \& Rodriguez 1994). We adopt the recent values $i=66^\circ$ and ${\cal D}=$11 kpc
(Fender et al. 1999). 

GRS 1915+105 stays almost permanently in a soft spectral state, although it has strong variability
of both flux and energy spectra. {\rxte} data have been extensively studied by Belloni et al. (2000).
The lightcurves are divided into 12 classes according to the color-color diagrams and the count rates.
Each class can be decomposed into only 3 basic spectral states: A, B and C.
Sobolewska \& \.{Z}ycki (2003) attempted the modeling of these three states, using the data of PCA
and HEXTE onboard {\rxte}. In the framework of MCD, they found that the X-ray spectra of A and B
(corresponding to the relatively higher luminosity) are complex: only one Comptonized disk plus another
model such as Comptonized disk, blackbody or Kerr disk can give an acceptable fitting.
They derived the accretion rate $\dot{M}=7.5\times 10^{18}$ g/s, namely $\dot{m}\sim 0.3$ for
the spectral state B. We choose the same observation data to do fitting with the present model.
For a comparison, we employ the absorption cross section given by Morrison \& McCammon (1983) to
correct the spectrum from our slim disk. The effective hydrogen column density is taken as
6.2$\times10^{22}$ $\mathrm{cm}^{-2}$ (Sobolewska \& \.{Z}ycki 2003). 
The spectral energy distribution is shown in Fig. 13. The dotted line is the spectrum from the
slim disk described in this paper. The Kerr MCD spectral component of Sobolewska \& \.{Z}ycki (2003)
is also shown by the grey dashed line. The hard tail is modeled by the thermal Comptonized
spectrum (dashed line) from the corona above the slim disk. With our slim disk model,
the best fitted parameters are $\alpha=0.01$, $\dot{m}=2.1$ and $f=0.03$.
It can be seen that the theoretical spectrum coincides with the observation very well in the
whole X-ray band. With these parameters, the transition radius is $\sim3.4$ $R_\mathrm{g}$.
In this case, the difference from the classical slim disk is not distinguishable.
But it is clear that our model is more self-consistent in the inner disk region.

\figurenum{13}
\centerline{\includegraphics[angle=-90.0,width=8.5cm]{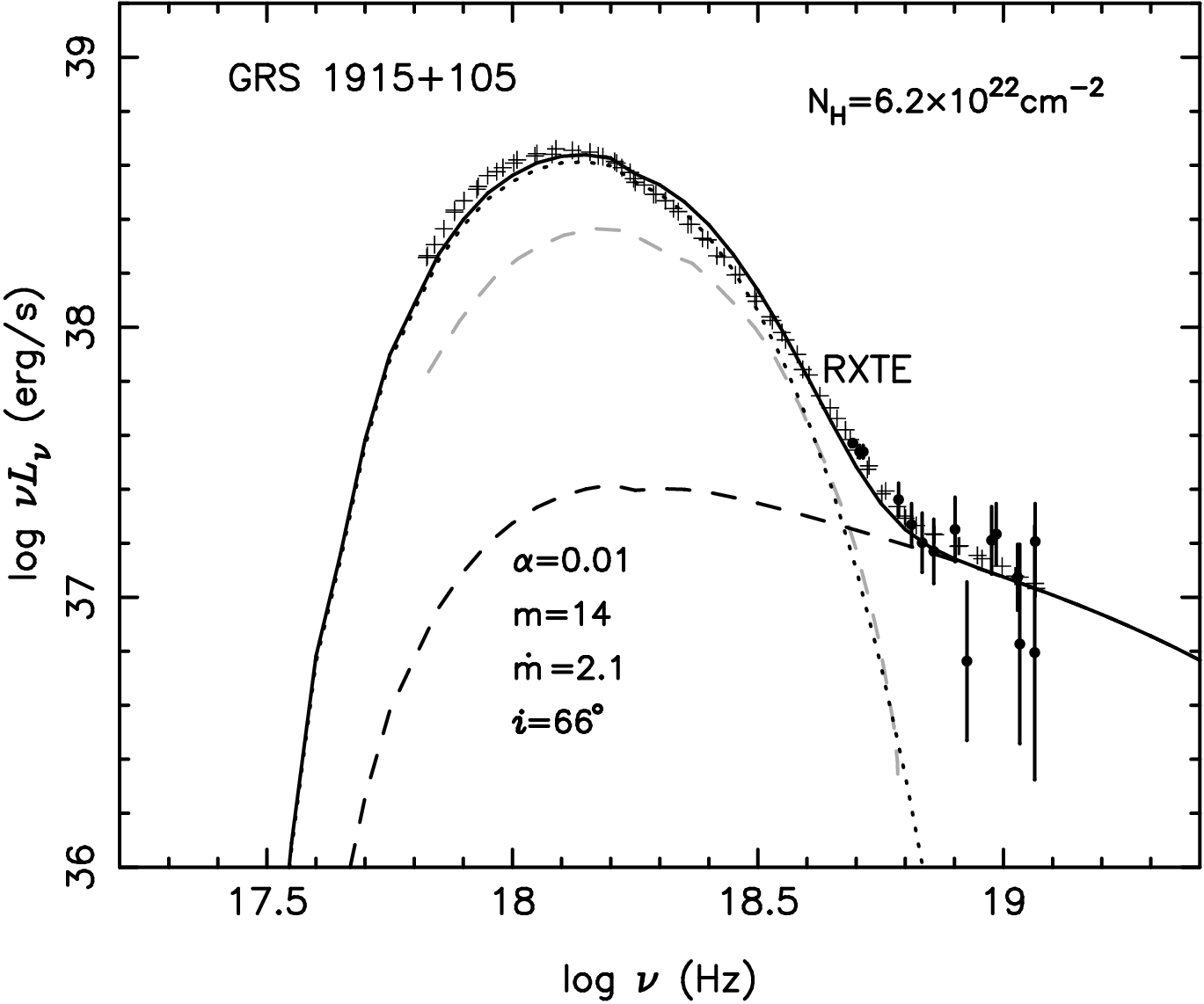}}
\figcaption{X-ray spectra of GRS 1915+105 after an absorption by the hydrogen column density of
6.2$\times\/10^{22}$cm$^{-2}$. Solid line is the fitting spectrum with our transition slim disk
(dotted line) plus the corona (dashed line). The {\rxte} observation data and the multi-color
Kerr metric disk spectrum (grey dashed line) are taken from Sobolewska \& \.{Z}ycki (2003).}
\label{grs1915}
\vglue 0.2cm

Since our slim disk covers rather wide ranges of accretion rate and viscosity parameter, it
can be also applied to other stellar black holes, such as galactic black hole binaries, superluminal
jet sources (microquasars) and ULXs. We have also successfully fitted the X-ray spectra of GRO J1655-40
(Tomsick et al. 1999), LMC X-1 (Gierli\'{n}ski et al. 2001). Their accretion rates are $\sim 0.5$
$\dot{M}_{\rm C}$. Ebisawa et al. (2003) applied the slim disk spectra to IC 342 source 1
(see also Waterai et al. 2001 for other ULXs). Since the basic quantities of ULXs,
e.g. the mass and the distance, are unknown, the spectral fitting is a little difficult. Once we
implement the present model into XSPEC, it will be possible to
model many other sources, especially XTE J1550-56 and IC 342 source 1 suspected having
super-critical accretion rates. This will help us understand more physics in ULXs. Making use of
the wide validity of the present model, we can understand the process of state transition, of which
the transition to an optically thin region exists.

\subsection{Narrow line Seyfert 1 galaxies}
Narrow line Seyfert 1 galaxies (NLS1s), as a special sub-group of Seyfert 1 galaxies,
are characterized by their unusual properties:
the H$\beta$ line is narrow (FWHM H$\beta<2000$ km/s) and relatively weak (less than a third
of the intensity of the [OIII]$\lambda 5007$ line), and the optical FeII lines are very strong.
%These characteristics locate NLS1s near the extreme value of the Boroson \& Green (1992)
%Eigenvector 1. 
It is revealed that the excess and the tendency of the spectrum to flatten
at low energy, have been found in several NLS1s, e.g. PKS 0558-504 (O'Brien et al. 2001),
Mrk 766 (Boller et al. 2001), PG 1244+026 (Fiore et al. 1998), PG 1404+226, PG 1440+356,
PG 1211+143 (George et al. 2000), IRAS 13224-3809 (Vaughan et al. 1999), 1H 0707-495
(Boller et al. 2002, Dewangan et al. 2002), RX J1702.5+3247 (Gliozzi et al. 2001) and
Ton S180 (Turner et al. 2001a, 2002). {\sax} and {\asca} observations of Akn 564 and
RE J1034+396 clearly show these extreme properties including a soft X-ray hump with a very
flat spectrum ($\nu F_{\nu}=const.$; Comastri et al. 2001; Puchnarewicz et al. 2001;
Turner et al. 2001b; Vignali et al. 2003; Romano et al. 2004). Other unusual X-ray properties
are discussed in Brandt et al. (1994); Comastri et al. (2001); Pounds, Done, \& Osborne (1995)
and Collinge et al. (2001). Earlier suggestions attributing the soft excess to blends of
emission lines are not supported by recent {\asca}, {\em XMM-Newton} and {\chandra}
observations (e.g. Turner, George, \& Netzer 1999; Puchnarewicz et al. 2001;
Turner et al. 2001a, 2002b; Collinge et al. 2001). {\chandra} shows that there is
not any feature of emission lines in soft X-ray band (Collier et al. 2001).
The model of less massive black hole with high accretion rate is preferred for its very
prominent soft X-ray hump (Boller et al. 1998; Wang \& Netzer 2003; Kawaguchi 2003).
In fact the feature of soft X-ray hump $\nu F_{\nu}=const$ is a natural explanation
of super-critical accretion disk (Wang \& Netzer 2003). With the black hole mass
estimated by the reverberation relation, Wang \& Netzer (2003) found half of NLS1s have
$L/L_{\rm Edd}>1$ in the sample of Ve\'ron-Cetty (2001). We believe that NLS1s are good candidates
for the applications of the present model.
 
As an application of the present model, we make an attempt to explain the observed soft X-ray
humps of Akn 564 and RE J1034+396. RE J1034+396 shows a very clear appearance
of flattening of soft X-ray spectrum from $\log \nu=16.5$ to $17.2$ in the $\nu L_{\nu}$
plot (Comastri et al. 2001; Puchnarewicz et al. 2001; Collinge et al. 2001; Turner et al. 2001).
%Such a component is first found in AGNs and quasars. 
Here we suggest that this feature
reflects the power of the slim disk in the two objects.

{\em RE J1034+396:}
The ultra-soft X-ray radiation of RE J1034+396 
was first discovered by Puchnarewicz et al. (1995). The spectrum shown in Fig. 14 shows a  clear
soft X-ray hump extending from $\log \nu=16.5$ (the lowest observed frequency) to $17.2$.
The data were taken from Puchnarewicz et al. (2001). They explained the strong soft X-ray spectrum 
of RE J1034+396 based on the standard accretion disk model (Czerny \& Elvis 1987). The resulting
accretion rate is $\dotm \sim 0.3-0.7$, which beyonds the scope of the standard disk model.
They also suggested a high inclination $i_{\rm obs}=60-75^{\circ}$ in order to lower the
luminosity.

The upper panel of Fig. 14 shows the comparison of our model with the observation
data. The FWHM of H$\beta$=2000 km/s and the luminosity $L_{5100}=10^{44}$ erg/s at 5100 \AA,
the mass of the black hole is estimated as $M_{\rm BH}=2.25\times 10^6M_{\odot}$ by the empirical
reverberation mapping relation (Kaspi et al. 2000). We assume the inclination angle $i=0^\circ$,
since no accurate estimation has been made in the literature.
The viscosity $\alpha=0.2$ is required from the fitting. We obtain the accretion rate $\dot{m}=1.2$. 
The factor $f=0.07$ is found, which is within the limit of equation 
(15). This result is in a rough agreement with that of self-similar solution
in Wang \& Netzer (2003).

\figurenum{14}
\centerline{\includegraphics[angle=-90.0,width=8.5cm]{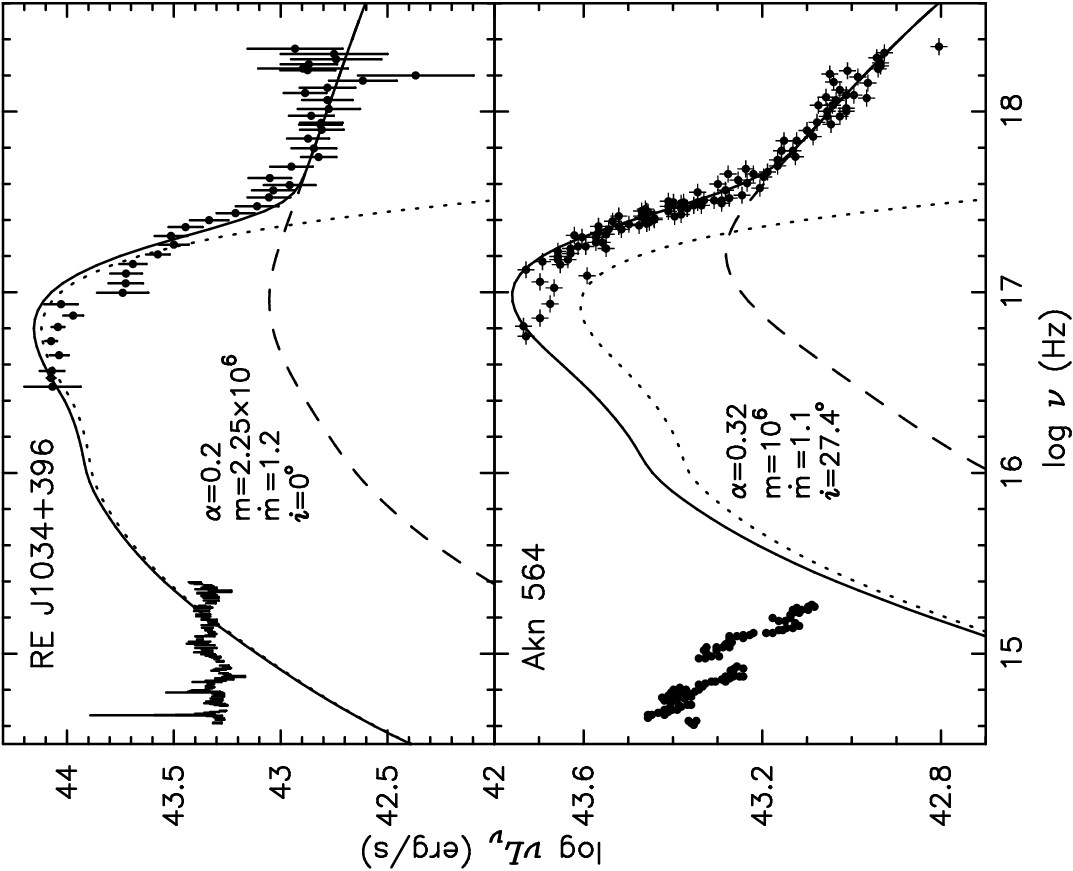}}
\figcaption{Broad-band spectra of REJ 1034+396 (top) and Akn 564 (bottom). The X-ray spectra are
both observated with {\sax}, and the data are taken from Puchnarewicz et al. (2001) and Comastri
et al. (2001). The dotted lines are the slim disk spectra. The hard tail is modeled by the
Comptonized spectra (dashed lines) from the corona above the slim disk, using our self-consistent
disk/corona model. The transition radii to the optically thin regions are 11.5 and 15.5 $R_\mathrm{g}$
for RE J1034+396 and Akn 564, respectively.}
\label{nls1}
\vglue 0.2cm

{\it Arakelian 564:}
Pounds et al. (2001) estimated that the black hole mass in Akn 564 is $\sim 10^7$ $M_{\odot}$
and the accretion rate $\dot{m}\approx 0.2-1$ from the break frequency of the power spectrum
of the light curves. The multiwavelength spectra of Akn 564 have been extensively studied by
Romano et al. (2004). They found that $M_{\rm BH}=4\times 10^6M_\odot$ and $\dotm \approx 1$.
Wang \& Netzer (2003) provided a simple method to estimate the black hole mass if the accretion
rate is super-critical, 
$M_{\rm BH}=2.8\times 10^6\left(\nu L_{\nu}/10^{44}{\rm erg/s}\right)M_{\odot}$,
which yields the black hole mass $2.0\times 10^6$ $M_{\odot}$
(Romano et al. 2004) for Akn 564. The current data has a few factors of uncertainty of $M_{\rm BH}$.
The inclination angle $i=27.4^\circ$ (Ballantyne et al. 2001), which was derived from the fitting
{\asca} spectra with the ionized reflection disk models of Ross \& Fabian (1993). The observed
spectral energy distribution of
Akn 564 is taken from Comastri et al. (2001). The comparison of the present model with the
observations provides $M_{\rm BH}=10^6M_{\odot}$, $\dot{m}=1.1$, $\alpha=0.32$ and $f=0.021$.
The hot corona is relatively weaker than that in RE J1034+396.

The fitting results indicate that the two NLS1s are moderate super-critical accreting sources.
In this case, the transition region plays great role in the emergent spectrum. The transition
radii, where $\tau_\mathrm{eff}=1$, are $\sim 11.5$ and $15.5R_\mathrm{g}$ for RE J1034+396 and
Akn 564, respectively. %In the vicinity of this transition radius, $\tau_\mathrm{eff}$ varies
%gradually so that a large region has $\tau_\mathrm{eff}\sim1$
%(see left panels of Fig. 5). 
This is just the main X-ray emitting zone.
Therefore, although the transition region influences little S-shaped curves (Fig. 4), it
has important effect on the spectrum from the disk. Our slim disk model provides more accurate
treatment about this transition region, so it has a wider application field than the previous slim
disk model. As stated by Pounds \& Reeves (2002), the soft X-ray humps are still poorly understood since
we do not know what drives the shape of soft X-ray spectrum. We need more observations.
The future work will focus on the spectral fitting in a sample so that we can discover
the primary parameters controlling the shape of soft X-ray spectrum.
%We found there are many NLS1s and quasars locate near the Eddington luminosity limit,
%the spectral fitting with the present model, as done in this paper, is expected in future.}

It is noted that the mass accretion rates for the three objects, GRS 1915+105, RE J1034+396
and Akn 564, are of $1\sim2$ critical accretion rates. For this range of accretion rate, the
slim disk with $\alpha$ viscous law have a large thermally unstable region.
% and viscously unstable regions, besides the transition regions for effective optical depths. 
These unstable regions are responsible for
the observed rapid and giant variabilities from microquasars and NLS1s. To describe the structures
accurately, the time-dependent slim disk model with transition regions is needed to explore
the variabilities of these super-critical accretors. We leave this topic in a future research.

\section{Conclusions}
The detailed structure and spectral calculations of the slim disk with transition region
(the effective optical depth $\tau_{\rm eff}\lesssim1$) have been presented. We have shown that
there is a quite large transition region in the slim disk. This region becomes larger with
the increase of viscosity parameter. We get an empirical formula for the transition radius, i.e.
$R_\mathrm{tr}/R_\mathrm{g}=95.38\alpha^{0.91}\dot{m}^{0.96}\exp\left(-0.1\dot{m}^{0.9}\right)$.
The maximum transition radius reaches $\sim 50$ $R_g$ at $\dot{m}\sim 15$ for a fixed
viscosity $\alpha$. In the slim disk regime, the transition radius
is $R_{\rm tr}/R_g\propto \alpha^{0.91}$ for a fixed accretion rate (e.g. $\dot{m}=10$) and
$\alpha>0.01$. The transition region affects not only the global disk structure, but also the
emergent spectrum. The transonic location moves outward and the local radiated spectrum departs
from blackbody radiation, because of the presence of the transition region. In comparison with
the classical slim disk, we have made more accurate treatment with the optically thin region,
in the calculations of both structure and spectrum. The present model covers a rather wide
parameter space. We have also shown how the soft X-ray spectrum changes from sub-critical, moderate
super-critical to the extreme super-critical accretion rate. We calculate the Comptonization
parameter for the slim disk, and show that the Comptonization is very essential for stellar-mass
black holes and less massive black holes ($\le 10^{6-7}$ $M_{\odot}$), whereas it is not very
important for more massive black holes with the masses larger than $10^9$ $M_{\odot}$.
This shows that the Comptonization process play a key role in the formation of spectra
in luminous objects. The present model can be widely applied to microquasars, ultra-luminous
X-ray sources and narrow line Seyfert 1 galaxies.

\acknowledgments
The authors thank the helpful discussions with J.F. Lu, M. Wu. Detailed reading of the
manuscript and productive comments from the refree, Ewa Szuszkiewicz, are also appreciated.
JMW thanks the support from National Science Foundation of China and Special Funds for
Major State Basic Research.

%\appendix

\clearpage

\end{document}